\documentclass[prd,aps,twocolumn,preprintnumbers,showpacs, nofootinbib,superscriptaddress,notitlepage]{revtex4-2}

\usepackage{slashed}
\usepackage{graphicx,color}
\usepackage{epsfig}
\usepackage{subfigure}
\usepackage{multirow}
\usepackage{epstopdf}
\usepackage{dcolumn}
\usepackage{bm}
\usepackage{color}
\usepackage{ulem}
\usepackage{enumitem}
\usepackage[T1]{fontenc}
\usepackage[colorlinks,
linkcolor=black,
anchorcolor=black,
citecolor=black
]{hyperref}

\def\bea{\begin{eqnarray}}
\def\eea{\end{eqnarray}}

\begin{document}

	\baselineskip=15pt
	
	\preprint{CTPU-PTC-25-03}

\title{ 
Probing ALP couplings to electroweak gauge bosons
}

\affiliation{ Department of Physics and Institute of Theoretical Physics, Nanjing Normal University, Nanjing, Jiangsu 210023, China}
\affiliation{Particle Theory and Cosmology Group, Center for Theoretical Physics of the Universe, Institute for Basic Science (IBS), Daejeon 34126, Korea }

\author{Jin Sun$^{2}$}
\email{sunjin0810@ibs.re.kr}
\author{Zhi-Peng Xing$^{1}$}
\email{zpxing@nnu.edu.cn(corresponding author)}
\author{Seokhoon Yun$^{2}$}
\email{seokhoon.yun@ibs.re.kr}

\begin{abstract}

Motivated by the more and more abundant experimental data, we revisit the couplings of 
axion-like particle (ALP) to electroweak gauge bosons across the ALP mass range from MeV to 100 GeV.
The  current and future experimental limits  on the couplings are extended. 
The ALP coupling to $W$-bosons gives rise to flavor-changing ALP-quark couplings at the one-loop level. 
These flavor-changing couplings deserve further investigation under current experimental constraints, especially those stemming from rare meson decays and neutral meson mixing processes.
Additionally, flavor-conserving couplings of the ALP to Standard Model (SM) fermions arise at the one-loop level as well from ALP-electroweak gauge boson couplings, even in the absence of tree-level couplings to these SM fermions, with consequent ALP decays to the SM fermions leading to constraints on the ALP-electroweak gauge boson couplings.
We also investigate processes relevant to 
$Z$-boson measurements, such as the invisible decay 
$Z\to a\gamma$, subsequent decays 
 $Z\to 3\gamma$ and $Z\to \gamma ll$, as well as constraints from oblique parameters  ($S,\, T,\, U$).
Our study  highlights that rare two-body decays of pseudoscalar mesons offer the most sensitive probes of ALP couplings to electroweak gauge bosons from the loop-induced flavor-violating interactions for ALP masses below the kinematic threshold, while 
$Z$-boson decays complementarily explore larger ALP masses.
Future lepton colliders, such as CEPC and FCC-ee operating at the 
$Z$-pole, along with SHiP, provide further opportunities to probe ALP couplings to electroweak gauge bosons.

\end{abstract}

\maketitle

\section{Introduction}

Probing new pseudoscalar particles with masses below the electroweak scale, predicted in some well-motivated extensions of the Standard Model (SM), plays an important role in particle physics.
A notable example is the quantum chromodynamics (QCD) axion, a pseudo-Nambu-Goldstone boson (PNGB) arising from spontaneous breaking of Peccei-Quinn symmetry, which is introduced to address the Strong CP problem in QCD \cite{Peccei:1977hh,Weinberg:1977ma,Wilczek:1977pj,Cheng:1987gp}.
 The QCD axion is characterized by a defining property: the linear proportionality between its mass and couplings, see e.g.~\cite{Kim:2008hd,DiLuzio:2020wdo}.
Axion-like particles (ALPs), on the other hand, generalize this concept, while they can be realized in a variety of new physics scenarios, see e.g.~\cite{Choi:2020rgn,Arias-Aragon:2022iwl} for a review. Unlike the QCD axion, ALPs do not have a mass-coupling proportionality, which opens up a significantly larger parameter space for their masses and couplings.
This freedom enables a rich phenomenology, investigated in both low-energy~\cite{Jaeckel:2010ni} and high-energy experiments~\cite{Beacham:2019nyx,Calibbi:2022izs}.
The PNGB nature of ALPs gives rise to interactions with SM particles respecting an approximate shift symmetry $a \to a + c$, which restricts their couplings to fermions in derivative forms and to gauge bosons via anomalous interactions within the framework of Effective Field Theory (EFT).
The leading-order interactions between ALPs and SM particles emerge at the dimension-five operator level, suppressed by the Peccei-Quinn breaking scale 
$f_a$. 
 For further details, see reviews in Refs.~\cite{Bauer:2017ris,Bauer:2021mvw}.
Due to their tiny couplings, the QCD axion and ALPs are also compelling candidates for dark matter
~\cite{Preskill:1982cy,Abbott:1982af,Dine:1982ah}.

Experimental searches for axions are primarily focusing on their interactions with photons, parameterized by the coupling constant $g_{a\gamma\gamma}$~\cite{Ringwald:2024uds}.
In the presence of a magnetic field, axions can convert into photons and vice versa via the Sikivie mechanism~\cite{Sikivie:1983ip} and the Primakoff effect~\cite{Primakoff:1951iae}.
Different axion mass ranges require different experimental approaches. Sub-MeV masses are investigated using laser-based experiments, helioscopes, and astrophysical observations, while the MeV–GeV mass range is primarily probed through beam dump experiments and high energy colliders.
Recently, several experiments have set new constraints on $g_{a\gamma\gamma}$. They include IBS-CAPP MAX~\cite{CAPP:2024dtx}, Oscillating Resonant Group AxioN (ORGAN)~\cite{Quiskamp:2023ehr}, Any Light Particle Search II (ALPS II)~\cite{Wei:2024fkf}, CERN Axion Solar Telescope (CAST)~\cite{CAST:2024eil}, and NEON~\cite{Park:2024omu}.
The photon-fusion production of ALPs, $\gamma\gamma \to a \to \gamma\gamma$
 (light-by-light scattering), has also been analyzed at $e^+e^-$ colliders such as FCC and ILC~\cite{RebelloTeles:2023uig}.
In addition to these experimental advancements, numerous new strategies have been proposed for ALP searches~\cite{Irastorza:2018dyq}.
These diverse experimental facilities have placed stringent bounds on
$g_{a\gamma\gamma}$. 

In addition to interactions with photons, 
 ALPs can interact with gluons~\cite{Chakraborty:2021wda,Bisht:2024hbs}, electroweak gauge bosons ($W^\pm$ and $Z$)~\cite{Izaguirre:2016dfi,Alonso-Alvarez:2018irt}, and fermions~\cite{DallaValleGarcia:2023xhh}.
While significant attention has been devoted to flavor-conserving ALP interactions, several mechanisms have been proposed to construct flavor-changing ALP interactions~\cite{Freytsis:2009ct, Celis:2014iua, Izaguirre:2016dfi, Choi:2017gpf, Heeck:2017wgr, Heeck:2019guh, MartinCamalich:2020dfe, Cheng:2020rla, Sun:2021jpw}.
This highlights the need to explore a broader range of theoretical frameworks and experimental opportunities in searches for ALPs. 
In particular, as first pointed out in \cite{Izaguirre:2016dfi}, ALP interactions with $W^\pm$ bosons can induce flavor-changing neutral current (FCNC) processes, leading to potentially observable signatures at low energies such as signals in rare meson decay processes.

In recent years, increasingly precise predictions for meson FCNC decays have been provided~~\cite{Bouchard:2013eph,Horgan:2013hoa,Buras:2014fpa,Alonso:2015sja,Buras:2015qea,Blake:2016olu,Gao:2019lta,Wang:2019wee,Beneke:2020fot,Huang:2024uuw,Buras:2024ewl,Fedele:2024jfj,Tian:2024ubt,Huber:2024rbw}, and numerous discrepancies between the SM predictions and experimental observations have come out~\cite{PDG2024}.
 For example, Belle-II uses an integrated luminosity $362 \,\textrm{fb}^{-1}$ to measure $\textrm{Br}(B^+ \to K^+ \nu \bar \nu)=23\pm 5^{+5}_{-4}$, exceeding the SM prediction by $2.7\sigma$ significance~\cite{Belle-II:2023esi}.
Similarly, the NA62 collaboration~\cite{NA62:2021zjw, NA62:2024pjp} has reported precise measurements of the branching ratio $\textrm{Br}(K^+\to \pi^+ \nu\bar\nu)$, significantly improving upon the previous results from the E949 experiment~\cite{BNL-E949:2009dza}.
Including them, Table~\ref{tab:decay} summarizes recent results and refined experimental upper limits for meson decays from various experiments.
These new results motivate us to update and extend previous analyses on meson FCNC processes involving an ALP.

The aforementioned FCNC processes are highly suppressed in the SM due to the combined effect of loop suppression and the Glashow–Iliopoulos–Maiani (GIM) mechanism~\cite{Glashow:1970gm}. The FCNC processes in the SM occur only at higher orders via weak interactions involving the exchange of at least two gauge bosons.
Notable examples include penguin and box diagrams, as well as long-distance double charged-current processes, such as
 $B^+\to  \tau^+ (\to K^+ \bar \nu)\nu$~\cite{Belle-II:2023esi}.
Given that the final state includes a pair of invisible neutrinos, experimental signals could be mimicked by invisible ALPs in processes like $B^+\to K^+ a$. Thus discrepancies between SM predictions and experimental results could be explained by light ALPs, which would manifest as missing-energy signals if the ALP does not decay inside the detector.
Conversely, if the ALP decays inside the detector, the corresponding leptonic final states, such as $B^+\to K^+ ll$, can place constraints on ALP parameters.
Similarly, pure leptonic decays, such as $B_s\to ll$, must also be considered.
 Furthermore, neutral-meson mixing systems are also influenced by flavor-changing interactions, and their impacts will be included in our analysis.

In this paper, we provide an updated and extended analysis on the experimental limits of the ALP couplings to electroweak gauge bosons across the ALP mass range from MeV to 100 GeV.
As mentioned, the ALP-$W^\pm$ couplings give rise to flavor-changing ALP-quark interactions $q_i-q_j-a$ at the one-loop level.
The associated physical processes, such as rare meson decays and neutral meson mixing, are analyzed to determine excluded regions in the ALP parameter space with the state-of-the-art SM predictions and new experimental data.
Additionally, we examine the impact of 
$Z$-boson precision measurements, including processes such as $\Gamma_Z$, $Z\to \gamma a$, $Z\to 3\gamma,\gamma ll$, and the ALP contributions to the electroweak oblique parameters ($S, T, U$). In particular, we found that the process $Z\to3\gamma$ significantly constrains the ALP couplings to electroweak gauge bosons above GeV scale. 
Finally, we analyze future experimental sensitivities from lepton colliders (e.g., CEPC and FCC-ee) operating at the 
$Z$-pole and the proposed SHiP experiment.

In order to provide a comprehensive picture for the current and future experimental limits on ALP couplings to electroweak gauge bosons, we study four distinctive scenarios. Since there are two independent gauge-invariant ALP couplings to electroweak gauge bosons, namely the ALP-$SU(2)_L$ gauge boson coupling $g_{aW}$ and the ALP-hypercharge gauge boson coupling $g_{aB}$,
these four scenarios are characterized by different relative sizes of the ALP-photon coupling to the ALP-$W^\pm$ coupling:
i) the photophobic ALP \cite{Craig:2018kne}, which suppresses ALP-photon interaction at a UV scale by cancellation between $g_{aW}$ and $g_{aB}$; ii) the ALP with $g_{aB}=0$; and iii) the ALP with the same sign of $g_{aW}$ and $g_{aB}$ to enhance the ALP-photon coupling;
and iv) the ALP with $g_{aW}=0$.

The rest of the paper is organized as follows. 
In section \ref{sec2}, we discuss low energy effective couplings of an ALP induced from the ALP couplings to electroweak gauge bosons at a UV scale. Section \ref{sec3} presents a comprehensive list of phenomenological observables relevant for probing the ALP couplings to electroweak gauge bosons. In section \ref{sec4}, we show our results on current and future experimental limits on the ALP-electroweak gauge boson couplings. We give our conclusions in section \ref{sec5}.

\section{ALP interactions with electroweak gauge bosons} \label{sec2}

The general ALP couplings with gauge bosons are written as 
\begin{eqnarray}\label{EW}
 \mathcal{L}_{EW} =-\frac{g_{aW}}4aW_{\mu\nu}^a\tilde{W}^{a\mu\nu}
  -\frac{g_{aB}}4aB_{\mu\nu}\tilde{B}^{\mu\nu}\;,
\end{eqnarray}
 where $a=1,2,3$ represents the SU(2) index,   $W^{\mu\nu}$ ($B^{\mu\nu}$) means $SU(2)_L$ ($U(1)_Y$) gauge bosons. 
$\tilde W^{\mu\nu}(B^{\mu\nu})$  are the dual field strength tensors. 
After symmetry breaking, the fields W and B are transformed into the physical fields $\gamma$ and Z,
\begin{eqnarray}\label{SM}
B_\mu = c_W  A_\mu -s_W  Z_\mu\;,\;\;  W^3_\mu =  s_W  A_\mu +  c_W  Z_\mu\;,
\end{eqnarray}
here $c_W \equiv \cos \theta_W$ and $s_W \equiv \sin  \theta_W$ with $\theta_W$ being the weak Weinberg mixing angle. 
Adopting the above transformation in Eq.~(\ref{EW}), the ALP-gauge boson interactions can be written as
\begin{eqnarray}
  && -\frac{a}{4}(g_{a\gamma\gamma}F_{\mu\nu}\tilde F^{\mu\nu}+g_{a\gamma Z}F_{\mu\nu}\tilde Z^{\mu\nu}
   +g_{aZZ}Z_{\mu\nu}\tilde Z^{\mu\nu}\nonumber\\
   &&\qquad+g_{aWW}
   W^+_{\mu\nu}\tilde{W}^{+\mu\nu})\;, 
\end{eqnarray}
where $F_{\mu\nu}$, $W_{\mu\nu}$, and $Z_{\mu\nu}$ are the field strength tensors of the photon, $W^\pm$, and Z bosons, respectively. 
And  the coupling coefficients can be expressed in terms of $g_{aW}$ and $g_{aB}$
\begin{eqnarray}\label{agauge}
&&g_{a\gamma\gamma}=g_{aW}s_W^2+g_{aB}c_W^2\;,
\quad
   g_{a\gamma Z}=2c_Ws_W (g_{aW}-g_{aB})\;,\nonumber\\
   && g_{aZ Z}=g_{aW}c_W^2+g_{aB}s_W^2\;.\quad
\end{eqnarray}

\begin{figure}
\centering
\includegraphics[width=0.9\linewidth]{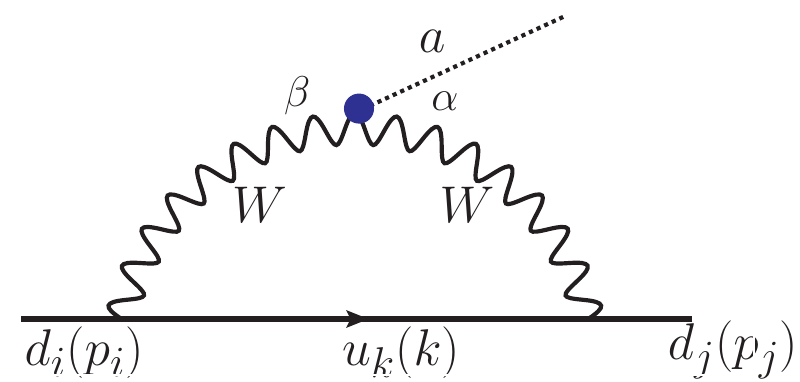}
\caption{\label{fig:abs}The flavor-changing $a-d_i-d_j$ interaction ($i\neq j$). Here the symbols in brackets mean the corresponding momentum.}
\end{figure}

In addition, the  ALP-$W^\pm$ couplings  arises the following interaction vertex 
\begin{eqnarray}\label{axion_quark}
    -i g_{aW}  p_{W\alpha}p_{W\beta}\epsilon^{\mu\nu\alpha\beta} a W_\mu W_\nu\;,
\end{eqnarray}
here $p_{W\alpha}$ and  $p_{W\beta}$ mean the four momentum of the W bosons.
The a-W-W interaction could contribute to the flavor-changing down-quark interaction as shown in Fig.~\ref{fig:abs}. 

Additionally, the SM charged vector-current interaction mediated by $W^\pm$ bosons are expressed by
\begin{eqnarray}\label{W_boson}
 \mathcal{L}_W&=&-\frac{g}{\sqrt{2}}(\bar u_L, \bar c_L,\bar t_L)\gamma^\mu V_{CKM}   
   \left( \begin{array}{c}
      d_L\\ s_L\\ b_L\end{array} \right)
  W^+_\mu +\mathrm{H.c.}.
\end{eqnarray}
Here $g$ means the gauge coupling constant of $SU(2)_L$ gauge group. And the subscript $L$ means the projection on the left.
And the CKM mixing matrix $V_{CKM}$ is parameterized by three rotation angles and one phase
\begin{eqnarray}
\left(\begin{array}{ccc}
c_{12}c_{13}&s_{12}c_{13}&s_{13}e^{-i\delta}\\-s_{12}c_{23}-c_{12}s_{23}s_{13}e^{i\delta}&c_{12}c_{23}-s_{12}s_{23}s_{13}e^{i\delta}&s_{23}c_{13}\\s_{12}s_{23}-c_{12}c_{23}s_{13}e^{i\delta}&-c_{12}s_{23}-s_{12}c_{23}s_{13}e^{i\delta}&c_{23}c_{13}
\end{array} \right),\nonumber\\
\end{eqnarray}
where the corresponding values are shown in Table.~\ref{tab:input}.
\begin{table}[htbp!]
\caption{The relevant parameters and input numbers.
}
\begin{tabular}{cccc}\hline\hline
 parameters  &input (GeV)   &parameters & input numbers  \\\hline
$m_Z$  &  91& $G_F$ & $1.1664\times 10^{-5}$GeV$^{-2}$
 \\\hline
$m_W$ & 80& $s_W^2$ & 0.23129\\\hline
$m_t$ & 173& $\Gamma_{B_s}$ & $65.81\times 10^{10}s^{-1}$ \\\hline
$m_b$ & 4.183& $\Gamma_{B^+}$ & $1/1638\times 10^{15}s^{-1}$ \\\hline
$m_c$ & 1.2730& $\Gamma_{B^0}$ & $1/1517\times 10^{15}s^{-1}$ \\\hline
$m_s$ & $0.0935$& $\Gamma_{K^+}$ & $1/1.2380\times 10^{8}s^{-1}$  \\\hline
$m_d$ & $0.0047$& $\Gamma_{K}$ & $1/5.116\times 10^{8}s^{-1}$ \\\hline
$m_u$ & $0.00216$&  & \\\hline
$m_\mu$ & $0.10566$&  & \\\hline
$m_{B_s}$ &  5.36693 & $\sin\theta_{12}$ &0.22501
 \\\hline
$m_{B^+}$ &  5.27941 & $\sin\theta_{13}$& 0.003732
 \\\hline
$m_{B^0}$ &  5.27972 & $\sin\theta_{23}$ &0.04183
\\\hline
$m_{K^+}$ &  $0.493677$ & $\delta$&1.147
 \\\hline
$m_{K_L}$ &  $0.497611$
& & \\\hline
$m_{\pi^+}$ &  $0.13957$
&   $f_K\sqrt{B_K}$& 0.132GeV\\\hline
$m_{\pi^0}$ &  $0.1349768$
&  $f_{B_d}\sqrt{B_d}$& 225MeV\\\hline
$m_{\rho^0}$ &  $0.770$
&  $f_{B_s}\sqrt{B_s}$ &274 MeV \\\hline
$m_{\phi}$ &  $1.020$
& & \\\hline
$m_{e}$ &  $0.511\times 10^{-3}$
& & \\\hline
$m_{\mu}$ &  $0.10566$
& & \\\hline
$m_{\tau}$ &  1.777
& & \\\hline
\end{tabular}
\label{tab:input}
\end{table}

Combing the interactions in Eqs.~(\ref{axion_quark},\ref{W_boson}), we can write down the amplitude in Fig.~\ref{fig:abs} as
\begin{eqnarray}
    iM&=&-g_{aW}\frac{g^2}{2}V_{CKM}^{ki}V_{CKM}^{kj*}\int \frac{d^4 k}{(2\pi)^4}  \bar u_j \gamma_\alpha (\slashed{k}-m_k)\gamma_\beta P_L u_{i}\nonumber\\
    &\times&
    \frac{(k-p_i)_\nu (k-p_j)_\mu \epsilon^{\mu\nu\alpha \beta}}{(k^2-m_k^2)((k-p_j)^2-m_W^2)((k-p_i)^2-m_W^2)}\;.
\end{eqnarray}
Here $i\neq j$ means the flavor-changing interactions.
At first glance, the amplitude should be UV divergences by analyzing the loop integral. After the dimensional regularization, the divergent term should be proportional to 
\begin{eqnarray}
div&=& g_{aW}\frac{g^2}{2}
V_{CKM}^{ki}V_{CKM}^{kj*}\frac{i}{4}\frac{\Gamma(2-d/2)}{16\pi^2}\bar u_j \gamma_\alpha \gamma^\gamma \gamma_\beta P_L u_i\nonumber\\
&\times&[(p_i)_\nu g_{\mu\gamma}+(p_f)_\mu g_{\nu\gamma}]\epsilon^{\mu\nu\alpha\beta},
\end{eqnarray}
where $d=4-\epsilon$. This divergence should be eliminated by the Renormalization. Before the renormalization, we can analyze the structure of divergent terms. We found that  the divergence term is independent of the quark mass. When summing over the up-type quarks (u,c,t), we naturally obtain the CKM matrix as
\begin{eqnarray}
    V_{CKM}^{ui}V_{CKM}^{uj*} +  V_{CKM}^{ci}V_{CKM}^{cj*} +
    V_{CKM}^{ti}V_{CKM}^{tj*}=0.
\end{eqnarray}
This shows that the divergences disappear due to the unitarity of CKM matrix when we consider the all three generation up-type quark in the propagator.  

Ignoring the quark mass of initial and final states and using $\epsilon^{\mu\nu\alpha\beta}\gamma_\nu \gamma_\alpha \gamma_\beta=6i\gamma^\mu\gamma^5$,  the finite term with $\bar u_j \slashed{p}_a P_L u_i$ can be obtained approximately as
\begin{eqnarray}
    &&2\int_0^1 dx \int_0^x dy \frac{6}{4(4\pi)^2}\log(1-y+\lambda y)\nonumber\\
    &&= \frac{3}{4(4\pi)^2}\left(-1+\frac{\lambda(1-\lambda+\lambda \log \lambda)}{(1-\lambda)^2}\right)\;,
\end{eqnarray}
here $\lambda=m_k^2/m_W^2$. Note that the first term (-1) will also disappear due to the CKM unitarity.
Therefore, combining the corresponding coefficients, we  obtain the following effective interaction as~\cite{Izaguirre:2016dfi}
\begin{eqnarray}\label{Eq:axion_interaction}
&&\mathcal{L}_{d_i\to d_j} \supset-g_{ad_id_j}(\partial_\mu a)\bar{d}_j\gamma^\mu\mathcal{P}_L d_i+\mathrm{H.c.}, \\
&& g_{ad_id_j} \equiv-\frac{3\sqrt{2}G_FM_W^2g_{aW}}{16\pi^2}\sum_{\alpha\in u,c,t}V_{\alpha i}V_{\alpha j}^*f(M_\alpha^2/m_W^2), \nonumber\\
&& f(x) \equiv\frac{x[1+x(\log x-1)]}{(1-x)^2}, \nonumber
\end{eqnarray}
here $G_F=1.1664\times 10^{-5}$GeV$^{-2}$ is the Fermi constant. $V_{ij}$ means the relevant Cabibbo-Kobayashi-Maskawa (CKM) matrix.
Note that for $x\ll1$, we obtain
\begin{eqnarray}
   \lim_{x\to 0} f(x)=x\;.
\end{eqnarray}
Note that the interaction is proportional to $M_\alpha^2/m_W^2$ for $M_\alpha<<m_W$.

Therefore, for the above flavor-changing couplings, the results is finite and only depends on the IR value of the effective coupling. Although the individual diagram in Fig.~\ref{fig:abs} are UV divergent, the divergences will cancel out when summed up intermediate up-type quark flavors $u_k$. This interesting feature is benefit from the two points: the unitarity of CKM matrix and quark-mass independent divergences.
This is in contrast with models possessing a direct ALP-quark coupling, in which the FCNC rate is sensitive to the UV completion~\cite{Batell:2009jf,Freytsis:2009ct}.

By further using the equation of motion, the above effective interaction for the on-shell fermions can be converted into
\begin{eqnarray}
   &&\mathcal{L}_{d_i\to d_j}
  = i g_{ad_id_j} a\bar{d}_j(m_{d_j}P_L-m_{d_i}P_R )  d_i+\mathrm{H.c.}.
\end{eqnarray}
This forms show that the main interactions should be RH chiral quark  structures due to $m_{d_i}>>m_{d_j}$.

\section{Phenomenological analysis} \label{sec3}

The above flavor-changing quark interaction induced by $a-W-W$ coupling contributes to different observables.
In this part, we drive the experimental constraints on the coupling $g_{aW}$ and ALP mass $m_a$,  as well as discussing possible ALP explanations for experimental anomalies.

\subsection{Meson FCNC decay}

The above quark couplings $a-d_i-d_j$ will mediate flavor-changing neutral current (FCNC) rare decays of heavy-flavor mesons at the tree level. 
The rare meson decays into mono-energetic final state mesons and on-shell ALP,  $M_1\to M_2 a$,   are the most sensitive probes of flavor-violating ALP couplings.
 This indicates that the relevant interactions are constrained by the corresponding physical processes as shown in Table.~\ref{tab:decay}.
\begin{table*}[htbp!]
\caption{The SM predictions and the experimental measurements of the meson decays. Upper limits are all given at 90\% confidence level (CL). Note that the symbol $*$ means the contribution is obtained by subtracting the tree-level effect from $B^+ \to \tau^+(\to K^{*+} \bar \nu)  \nu$, $[(10.86\pm 1.43)-(1.07\pm 0.10)]\times 10^{-6}$.}
\begin{tabular}{cccc}\hline\hline
quark transition&Observable  &SM prediction ($\times 10^{-6}$) &Experimental data ($\times 10^{-6}$) \\\hline
\multirow{6}{*}{$s\to d$}
&$Br(K^+ \to \pi^+ \nu \bar \nu)$ & $(8.42\pm 0.61)\times 10^{-5}$~\cite{Hou:2024vyw}&$(13.0^{+3.3}_{-3.0})\times 10^{-5} $(NA62~\cite{NA62:2021zjw,NA62:2024pjp})\cr\cline{2-4}
&$Br(K_L \to \pi^0 \nu \bar \nu)$ & $(3.41\pm 0.45)\times 10^{-5}$~\cite{Hou:2024vyw} &$<3\times 10^{-3}$(KOTO~\cite{KOTO:2018dsc})\cr\cline{2-4}
&$Br(K^+ \to \pi^+ ee)$ & $0.3\pm0.03$~\cite{DAmbrosio:1997eof} &$0.3\pm0.009$(NA48~\cite{PDG2024})\cr\cline{2-4}
&$Br(K^+ \to \pi^+ \mu\mu)$ & $(9.4\pm0.6)\times 10^{-2}$~\cite{DAmbrosio:1997eof} &$(9.17\pm0.14)\times 10^{-2}$(NA62~\cite{PDG2024})\cr\cline{2-4}
&$Br(K_L \to \pi^0 ee)$ & $(3.38\pm0.92)\times 10^{-5}$~\cite{Buras:2022qip} &$<2.8\times 10^{-4}$(KTEV~\cite{KTeV:2003sls})\cr\cline{2-4}
&$Br(K_L \to \pi^0 \mu\mu)$ & $(1.39\pm 0.27)\times 10^{-5}$~\cite{Buras:2022qip} &$<3.8\times 10^{-4}$(KTEV~\cite{KTEV:2000ngj})
\\\hline\hline
\multirow{12}{*}{$b\to d$}&$Br(B^+ \to \pi^+ \nu \bar \nu)$ & $0.140\pm0.018$~\cite{Hou:2024vyw}&$<14$(Belle II~\cite{Belle-II:2022cgf})\cr\cline{2-4}
&$Br(B^0 \to \pi^0 \nu \bar \nu)$ & $0.0652\pm 0.0085$~\cite{Hou:2024vyw}&$<9$(Belle II~\cite{Belle-II:2022cgf})\cr\cline{2-4}
&$Br(B^+ \to \rho^+ \nu \bar \nu)$ & $0.406\pm 0.079$~\cite{Hou:2024vyw}&$<30$(Belle II~\cite{Belle-II:2022cgf})\cr\cline{2-4}
&$Br(B^0 \to \rho^0 \nu \bar \nu)$ & $0.189\pm 0.036$~\cite{Hou:2024vyw}&$<40$(Belle II~\cite{Belle-II:2022cgf})\cr\cline{2-4}
&$Br(B^+ \to \pi^+ ee)$ & $ (1.95\pm0.61)\times 10^{-2}$~\cite{Wang:2012ab}
&$<5.4\times 10^{-2}$(Belle-II~\cite{Belle:2024cis})\cr\cline{2-4}
&$Br(B^0 \to \pi^0 ee)$ & $ (0.91\pm0.34)\times 10^{-2}$~\cite{Wang:2012ab}
&$<7.9\times 10^{-2}$(Belle-II~\cite{Belle:2024cis})\cr\cline{2-4}
&$Br(B^+ \to \rho^+ ee)$ & O(0.02)~\cite{PDG2024}
&$<0.467$(Belle-II~\cite{Belle:2024cis})\cr\cline{2-4}
&$Br(B^+ \to \pi^+ \mu\mu)$ & $ (1.95\pm0.61)\times 10^{-2}$~\cite{Wang:2012ab}
&$(1.78\pm0.23)\times 10^{-2}$(LHCb~\cite{PDG2024})\cr\cline{2-4}
&$Br(B^0 \to \pi^0 \mu\mu)$ & $ (0.91\pm0.34)\times 10^{-2}$~\cite{Wang:2012ab}
&$<5.9\times 10^{-2}$(Belle-II~\cite{Belle:2024cis})\cr\cline{2-4}
&$Br(B^+ \to \rho^+ \mu\mu)$ & O(0.02)~\cite{PDG2024}
&$<0.381$(Belle II~\cite{Belle:2024cis})\cr\cline{2-4}
&$Br(B \to  \mu \mu)$ & $(1.03\pm0.05)\times 10^{-4}$~\cite{Beneke:2019slt}&$<1.5\times 10^{-4}$(CMS~\cite{CMS:2022mgd})\cr\cline{2-4}
&$Br(B \to  \tau \tau)$ & 0.03~\cite{Sahoo:2012ynk}&$<2100$(CMS~\cite{CMS:2022mgd})
\\\hline\hline
\multirow{15}{*}{$b\to s$}&$Br(B^+ \to K^+ \nu \bar \nu)$ & 
$(4.97\pm0.37)$~(HPQCD\cite{Parrott:2022zte})
&$23\pm 5^{+5}_{-4}$(Belle II~\cite{Belle-II:2023esi})  \cr\cline{2-4}
&$Br(B^0 \to K^0 \nu \bar \nu)$ & $3.85\pm0.52$~\cite{Hou:2024vyw}&$<26$(Belle~\cite{Belle:2017oht})  \cr\cline{2-4}
&$Br(B^+ \to K^{*+} \nu \bar \nu)$ & $9.79\pm 1.43^*$~\cite{Becirevic:2023aov}&$<61$(Belle~\cite{Belle:2017oht} )\cr\cline{2-4}
&$Br(B^0 \to K^{*0} \nu \bar \nu)$ &$9.05\pm 1.37$~\cite{Becirevic:2023aov}&$<18$(Belle~\cite{Belle:2017oht}) \cr\cline{2-4}
&$Br(B_s \to \phi \nu \bar \nu)$  & $9.93\pm 0.72$~\cite{Hou:2024vyw}&$<5400$(LEP DELPHI~\cite{DELPHI:1996ohp}) \cr\cline{2-4}
&$Br(B^+ \to K^+ ee)$  & $0.191\pm0.015$~\cite{Buras:2022qip}&$0.56\pm 0.06$(Belle~\cite{PDG2024}) \cr\cline{2-4}
&$Br(B^0 \to K^0 ee)$  & $0.51\pm0.16$~\cite{Wang:2012ab}&$0.25\pm 0.11$(Belle~\cite{PDG2024}) \cr\cline{2-4}
&$Br(B \to K^* ee)$  & $0.239\pm0.028$~\cite{Buras:2022qip}&$1.42\pm 0.49$(Belle-II~\cite{Taniguchi:2024mls}) \cr\cline{2-4}
&$Br(B^+ \to K^+ \mu\mu)$  & $0.191\pm0.015$~\cite{Buras:2022qip}&$0.1242\pm0.0068$(CMS~\cite{CMS:2024syx}) \cr\cline{2-4}
&$Br(B^0 \to K^0 \mu\mu)$  & $0.51\pm0.16$~\cite{Wang:2012ab}&$0.339\pm0.035$(Belle~\cite{PDG2024}) \cr\cline{2-4}
&$Br(B \to K^* \mu\mu)$  & $0.239\pm0.028$~\cite{Buras:2022qip}&$1.19\pm0.32$(Belle-II~\cite{Taniguchi:2024mls}) \cr\cline{2-4}
&$Br(B_s \to \phi \mu\mu)$  & $(0.27\pm0.025)$~\cite{Buras:2022qip}&$0.814\pm0.047$(LHCb~\cite{LHCb:2021zwz}) \cr\cline{2-4}
&$Br(B^+ \to K^+ \tau\tau)$  & $0.12\pm0.032$~\cite{Wang:2012ab}&$<2250$(BaBar~\cite{BaBar:2016wgb}) \cr\cline{2-4}
&$Br(B_s \to  \mu\mu)$  & $(3.78\pm 0.15)\times 10^{-3}$~\cite{Buras:2022qip}&$(3.34\pm0.27)\times 10^{-3}$(CMS~\cite{CMS:2022mgd})\cr\cline{2-4}
&$Br(B_s \to  \tau\tau)$  & 0.8~\cite{Sahoo:2012ynk} &$<6800$(LHCb~\cite{LHCb:2017myy})
\\\hline
\hline
\end{tabular}
\label{tab:decay}
\end{table*}

We  exclusively focus on the bounds derived on the flavor-changing ALP couplings. For estimating the transition matrix element of  meson rare decays, we take B meson as example with the form factors  as~\cite{Ball:2004ye,Ball:2004rg}
\begin{eqnarray}
&&\langle P(p)|\bar q b|B(p_B)\rangle=\bigg(\frac{m_B^2-m_p^2}{m_b-m_q}\bigg)f^P_0(q^2),\nonumber\\
&&\langle V(p)|\bar q \gamma_5 b|B(p_B)\rangle=\frac{-i2m_V \epsilon^*\cdot q }{m_b+m_q} A_0(q^2).
\end{eqnarray}
For $B\to K^*a$ decay, the $K^*$ meson actually only have longitudinally polarization contribution since the ALP is a pseudoscalar particle.
For the Kaon rare decays, we can estimate its amplitude under the vertor current conserved assumption~\cite{Izaguirre:2016dfi}. 
Note that the matrix element for $K^0\to \pi^0 a$ is related to $K^\pm\to \pi^\pm a$ by isospin symmetry. Therefore, the matrix element for $K_L(K_S)$ mass eigenstate is obtained by taking the imaginary (real) part of $K^\pm\to \pi^\pm a$ matrix element~\cite{Marciano:1996wy}.
Therefore the corresponding decays for B and K mesons are expressed by
\begin{eqnarray}
\Gamma(B^{+}\to\pi^{+}a)&=&\frac{m_B^3}{64\pi}|g_{abd}|^2\left|f_0^{B\to\pi}\left(m_a^2\right)\right|^2\left(1-\frac{m_\pi^2}{m_B^2}\right)^2\nonumber\\
&\times&\lambda^{1/2}\left(\frac{m_\pi}{m_B},\frac{m_a}{m_B}\right), \nonumber\\
\Gamma\left(\bar{B}^0\to\pi^0a\right) &=&\frac{1}{2} \Gamma\left(B^-\to\pi^-a\right) ,\nonumber \\
 \Gamma(B\to Ka)&=& \frac{m_B^3}{64\pi}|g_{abs}|^2\left(1-\frac{m_K^2}{m_B^2}\right)^2|f^{B\to K}_0(m_a^2)|^2\nonumber\\
 &\times&\lambda^{1/2}\left(\frac{m_K}{m_B},\frac{m_a}{m_B}\right),\nonumber \\
\Gamma(B\to K^*a)&=& \frac{m_B^3}{64\pi}|g_{abs}|^2|A_0(m_a^2)|^2\lambda^{3/2}\left(\frac{m_{K^*}}{m_B},\frac{m_a}{m_B}\right), \nonumber\\
\Gamma(K^+\to\pi^+a)&=& \frac{m_{K^+}^3}{64\pi}\left(1-\frac{m_{\pi^+}^2}{m_{K^+}^2}\right)^2|g_{asd}|^2\nonumber\\
&\times&\lambda^{1/2}\left(\frac{m_{\pi^+}}{m_{K^+}},\frac{m_a}{m_{K^+}}\right), \nonumber\\
\Gamma(K_L\to\pi^0a) &=&\frac{m_{K_\mathrm{L}}^3}{64\pi}\left(1-\frac{m_{\pi^0}^2}{m_{K_\mathrm{L}}^2}\right)^2\mathrm{Im}(g_{asd})^2\nonumber\\
&\times&\lambda^{1/2}\left(\frac{m_{\pi^0}}{m_{K_L}},\frac{m_a}{m_{K_L}}\right), 
\end{eqnarray}
here $\lambda(x,y)=[1-(x+y)^2][1-(x-y)^2]$. 
Note that the factor $1/2$ in $\Gamma(\bar B^0\to \pi^0 a)$ comes from the quark components $\pi^0=(\bar u u+\bar d d)/\sqrt{2}$.
The other decay processes can be obtained by the corresponding transformations on the masses  and form factors.
Using light-cone sum rules~\cite{Ball:2004ye,Ball:2004rg},
the corresponding expressions are 
\begin{eqnarray}
   &&   A_0^{B\to K^*}(m_a^2)=\frac{1.364}{1-m_a^2/(5.28)^2}+ \frac{-0.990}{1-m_a^2/36.78}\;,\nonumber\\
    &&   A_0^{B\to \rho}(m_a^2)=\frac{1.527}{1-m_a^2/5.28^2}+ \frac{-1.220}{1-m_a^2/ 33.36}\;,\nonumber\\
  &&   A_0^{B_s\to \phi}(m_a^2)=\frac{3.310}{1-m_a^2/5.28^2}+ \frac{-2.835}{1-m_a^2/ 31.57},\nonumber\\
   && f_0^{B\to \pi}(m_a^2)=\frac{0.258}{1- 
    m_a^2/33.81}\;,\nonumber\\
  &&  f_0^{B\to K}(m_a^2)=\frac{0.330}{1-m_a^2/37.46}\;.
\end{eqnarray}
Note that the above forms fix the axion mass unit with $m_a/\text{GeV}$.
In all cases these couplings are renormalized at the scale of the measurement, but because the flavor-changing ALP couplings do not run below the weak scale, it is equivalent to use couplings renormalized at the weak scale. Additionally, we expect that subprocesses of the type $B^-\to \pi^-a$ via ALP-pion mixing give rise to subdominant contributions to the $B^-\to \pi^-$ rate. The assumtions can be applied into other meson decays.

We should stress that the NP effects in the $K^+$ and in the $K_L$ decay are in general highly correlated by  the Grossman-Nir bound~\cite{Grossman:1997sk} with
\begin{eqnarray}
    \frac{\mathrm{BR}(K_{L}\to\pi^{0}\nu\bar{\nu})}{\mathrm{BR}(K^{+}\to\pi^{+}\nu\bar{\nu})}<4.3\;.
\end{eqnarray}

\subsection{semi- and pure- leptonic meson decays}

Although ALPs do not interact with fermions at the UV scale, the one-loop RG evolution from the UV scale
 $\Lambda$ down to the weak scale
induces the flavor conserving coupling to fermions $\mathcal{L}\supset g_{aFF}\partial_\mu a \bar{F}\gamma^\mu\gamma^5F$~\cite{Craig:2018kne} as
\begin{eqnarray}
g_{aFF}&=&\frac{3\alpha^2 }{4}\left[\frac{3}{4s^4_W}\frac{g_{aW}}{g^2}+\frac{(Y_{F_L}^2+Y_{F_R}^2)}{c^4_W}\frac{g_{aB}}{g^{\prime 2}}\right]\log\frac{\Lambda^2}{m_W^2}\nonumber\\
&+&\frac{3}{2}Q_F^2 \frac{\alpha}{4\pi}g_{a\gamma\gamma}\log\frac{m_W^2}{m_F^2}\;,
\end{eqnarray}
where $Y_{F_{L,R}}$ are the hypercharges of the chiral fermion F fields, and $s_W=\sin \theta_W$ and $c_W=\cos\theta_W$. In our study, we choose $\Lambda=10$TeV.

The above forms $g_{a\gamma\gamma}$ in Eq.~(\ref{agauge}) comes from the tree level contribution.
The loop corrections also contributes to the coupling  given by
\begin{eqnarray}
\frac{g^{loop}_{a\gamma\gamma}}{e^2}&=&\frac{g_{aW}}{2\pi^2 }B_2\left(\frac{4m_W^2}{m_a^2}\right)-\sum_F\frac{N_c^FQ_F^2}{2\pi^2}g_{aFF}B_1\left(\frac{4m_F^2}{m_a^2}\right),\nonumber\\
\end{eqnarray}
with the form of the functions as
\begin{eqnarray}
 &&B_1(x)=1-xg(x)^2,\quad B_2(x)=1-(x-1)g(x)^2\;,\nonumber\\
 && g(x)=\left\{\begin{array}{ll}\arcsin\frac{1}{\sqrt{x}}\qquad   \qquad   \qquad \mbox{for}\qquad x\geq1 \\\\
    \frac{\pi}{2}+\frac{i}{2}\log\frac{1+\sqrt{1-x}}{1-\sqrt{1-x}},\qquad \mbox{for}\qquad x<1.\end{array}\right.
\end{eqnarray}
Here $N^F_c=1(3)$ stands for the charged leptons (quarks). 
Therefore, the total ALP couplings with photons should consider the tree and loop level contributions  simultaneously, $g^{eff}_{a\gamma\gamma}=g_{a\gamma\gamma}+g^{loop}_{a\gamma\gamma}$.

Correspondingly, the decay widths are obtained by 
\begin{eqnarray}
&&\Gamma_{total}=\sum_F \Gamma(a\to F\bar{F}) +\Gamma(a\to\mathrm{hadrons})+ \Gamma(a\to\gamma\gamma)\;,\nonumber\\
&&\Gamma(a\to F\bar{F})=\frac{N^F_c}{2\pi}m_am_F^2 g_{aFF}^2\left(1-\frac{4m_F^2}{m_a^2}\right)^{1/2}\;,\\
&&\Gamma(a\to\mathrm{hadrons})=\frac{1}{8\pi^3}\alpha_s^2m_a^3\left(1+\frac{83}{4}\frac{\alpha_s}{\pi}\right)\left|\sum_{q=u,d,s}g_{aqq}\right|^2\;,\nonumber\\
&&\Gamma(a\to\pi^a\pi^b\pi^0)=\frac{\pi}{24}\frac{m_am_\pi^4}{f_\pi^2}\left[\frac{g_{auu}-g_{add}}{32\pi^2}\right]^2g_{ab}\left(\frac{m_\pi^2}{m_a^2}\right),\nonumber\\
&&\Gamma(a\to\gamma\gamma)=\frac{m_a^3}{64\pi}|g^{eff}_{a\gamma\gamma}|^2\;.
\end{eqnarray}
Here the fermion F stands for the charged leptons $e,\mu,\tau$ and heavy quarks $c,b,t$.
Note that the decay channel $a\to 3\pi$ only applies for $m_a>3m_\pi$.
Here $f_\pi=0.13$GeV means the pion meson decay constant. 
For $m_a>1$GeV, the $a\to 3\pi$ channel will be absorbed into the $a\to $ hadrons one.
And the corresponding functions are defined by
\begin{eqnarray}
   g_{00}(r)&=&\frac{2}{(1-r)^2}\int_{4r}^{(1-\sqrt{r})^2}dz\sqrt{1-\frac{4r}{z}}\lambda^{1/2}(\sqrt{z},\sqrt{r}),\nonumber\\
   g_{+-}(r)&=&\frac{12}{(1-r)^2}\int_{4r}^{(1-\sqrt{r})^2}dz\sqrt{1-\frac{4r}{z}}\nonumber\\
   &\times& (z-r)^2
   \lambda^{1/2}(\sqrt{z},\sqrt{r}) \;.
\end{eqnarray}
 Furthermore, the branching ratios for the corresponding decay chains can be determined.
 Notably, the total decay width  $\Gamma_{total}$ plays a critical role in determining whether the ALP decays within the experimental detector.

If ALP can decay into the above SM particles within the length of detector, we should consider the ALP decay probability defined as
  \begin{eqnarray}
      \mathcal{P}_{\mathrm{dec}}^{a}=1-\exp\left(-\ell_D\Gamma_{a}\frac{M_{a}}{p_{a}}\right)\;.
  \end{eqnarray}
Here $\ell_D$ refers to the transverse radius size of the detector, with 2.5 meters for NA62 and KOTO, 1.5 meters for BaBar, 1.8 meters for NA48, 1 meter for KTeV, 2 (4) meters for Belle (Belle-II), 7.5 meters for CMS, 6 meters for LEP, and 5 meters for LHCb, respectively.

Therefore, the branching ratios  of the  subsequent semi-leptonic decays are
\begin{eqnarray}\label{alepton}
    Br(M_1\to M_2 ll)= Br(M_1\to M_2 a) Br(a\to ll)\mathcal{P}_{\mathrm{dec}}^{a}\;. \nonumber\\
\end{eqnarray}
The experimental data can constrain the model parameters for the decay $a\to ll$
 if it occurs kinematically, $m_a>2m_l$.

In addition to the aforementioned semi-leptonic decays, pure leptonic decays can serve as sensitive probes of flavor-changing ALP couplings. Properly accounting for their interference, we find that the ALP contribution modifies the branching ratios~\cite{Bauer:2021mvw}
\begin{eqnarray}
R(B_{s,d})&=&\frac{\mathrm{Br}(B_{s,d}\to ll)}{\mathrm{Br}(B_{s,d}\to  ll)_{\mathrm{SM}}}\\
&=&\left|1-\frac{g_{all}}{C_{10}^{\mathrm{SM}}(\mu_b)}\frac{\pi}{\alpha(\mu_b)}\frac{v_{SM}^2}{1-m_a^2/m_{B_{s,d}}^2}\frac{g_{abs,abd}}{V_{ts}^*V_{tb}}\right|^2\nonumber,
\end{eqnarray}
here $C_{10}^{SM}(m_b)\approx -4.2$ means the wilson coefficient of the operator $O_{10}=\bar{s}_{L}\gamma_{\mu}b_{L}\bar{\ell}\gamma^{\mu}\gamma_{5}\ell$~\cite{Hiller:2014yaa}.
We choose the strongest bounds from muons as shown in Table.~\ref{tab:decay}.

\subsection{Neutral-meson mixing}

The meson mixing $\Delta F=2$ can also place bounds on the model parameters.
Neutral meson mixing is governed by the off-diagonal entries of the two-state Hamiltonian $\hat H=\hat M-i\hat \Gamma/2$, where the Hermitian $2\times 2$ matrices $\hat M$ and $\hat \Gamma$ describe the off-shell and on-shell transitions respectively. 
The effective Hamiltonian $H^{ij}_{\Delta F=2}$ receives contributions from both the SM and NP effects, with  
 mixing amplitude  defined as
\begin{eqnarray}
\mathcal{L}_{\rm{eff}}&=&\bar{q}_1\Gamma_1q_2\bar{q}_1\Gamma_2q_2 
\quad \Rightarrow\quad M_{12}=\frac{1}{2m_P}\langle\bar{P}|\mathcal{H}_{\mathrm{eff}}|P\rangle\nonumber\\
&=&-\frac1{2m_P}\langle\bar{P}|\bar{q}_1\Gamma_1q_2\bar{q}_1\Gamma_2q_2|P\rangle\;,
\end{eqnarray}
here $q_{1,2}=ds,db,sb$ correspond to the mixing of $K^0-\bar K^0$,   $B_d-\bar B_d$ and  $B_s-\bar B_s$, respectively. 
And $\Gamma_{1,2}$ means the different interaction structure combination.
When considering the specific meson-mixing system, the mass differences for the K, $B_d$, $B_s$ are 
\begin{eqnarray}
 &&   \Delta m_K= 2\Re(M_{12}^K),\;
   \Delta m_{B_q}= 2|M_{12}^{ib}|\;.
\end{eqnarray}
Note that for $K-\bar K$ mixing system, we need modify the absolute value as the real part.

The hadronic matrix elements of the relevant operators  can then be written in terms of hadronic parameters $B_P(\mu)$ as
\begin{eqnarray}
\frac1{m_P}\langle\bar{P}|\bar{q}_1\Gamma_1q_2\bar{q}_1\Gamma_2q_2|P\rangle = f_P^2 m_P \eta(\mu) B_P(\mu)\;,
\end{eqnarray}
where the decay constant of $f_P$ meson and  the bag parameter $B_P$ can refer to Refs.~\cite{Lenz:2010gu,Carrasco:2015pra} with numbers  shown in Table.~\ref{tab:input}.
The meson $P(H\bar q)$ is composed of one heavy quark H and one light antiquark $\bar q$.

For the above interaction in Eq.~(\ref{Eq:axion_interaction}), it can  naturally lead to the four-fermion operators. Adopting the notations in Ref.~\cite{Ciuchini:1998ix}, the  relevant
two effector operators are expressed in the following
\begin{eqnarray}
&&\mathcal{H}_{eff}= \tilde c_2(\mu_H)\tilde O_2+\tilde c_3(\mu_H)\tilde O_3,\nonumber\\
&& \tilde O_2=\bar q_L^\alpha  H_R^\alpha \bar q_L^\beta  H_R^\beta\;,  \quad 
 \tilde O_3=\bar q_L^\alpha  H_R^\beta \bar q_L^\beta  H_R^\alpha\;.
\end{eqnarray}
For the case of $m_a\leq m_b$, the wilson coefficients are
\begin{eqnarray}
    &&\tilde c_2(\mu_H)=-\frac{m_H^2(\mu_H)}{2}\frac{N_c^2 A_+ -A_-}{(N_c^2-1)}g_{a d_i d_j}^2\;,\nonumber\\
    &&\tilde c_3(\mu_H)=-\frac{m_H^2(\mu_H)}{2}\frac{N_c( A_- -A_+)}{(N_c^2-1)}g_{a d_i d_j}^2\;,
\end{eqnarray}
here $A_\pm=1/[(m_H\pm (m_{P}-m_H))^2-m_a^2]$, and $N_c=3$ means the color numbers for quarks.
The superscripts $\alpha,\beta$ mean the corresponding color index.
The normalization factors $\eta(\mu)$ are conventionally obtained using the naive vacuum insertion (VIA) approximation for the matrix elements. Under the VIA condition, the above two operators  $\tilde O_{2,3}$ will lead to $\eta(\mu)$ as
\begin{eqnarray}
 && \tilde \eta_2(\mu_H)=-\frac{1}{2}\left(1-\frac{1}{2N_c}\right) \left(\frac{m_{P}}{m_H+m_q}\right)^2\;,\nonumber\\
 &&  \tilde \eta_3(\mu_H)=-\frac{1}{2}\left(\frac{1}{N_c}-\frac{1}{2}\right) \left(\frac{m_{P}}{m_H+m_q}\right)^2\;.
\end{eqnarray}

For the case $m_a>m_H$, the relevant Wilson coefficients are replaced by
\begin{eqnarray}
&&\tilde c_2(\mu_b)=\left(0.983\eta^{-2.42}+0.017\eta^{2.75}\right)\tilde c_2(\mu_a),\\
&&\tilde{c}_2(\mu_a)=\frac{m_b^2(\mu_a)}{2m_a^2}\left(g_{a d_i d_j }\right)^2, \eta=[\alpha_s(\mu_a)/\alpha_s(\mu_b)]^{6/23}\;,\nonumber
\end{eqnarray}
here we consider the running effects from axion mass scale  down to the scale $m_b$.

These off-diagonal matrix elements are directly related to the experimentally measured quantities as shown in Table.~\ref{tab:mixing}.
We found that  the SM predictions on $\Delta m_F$ are consistent with experimental data within errors and the uncertainties from theoretical non-perturbative QCD effects are larger than those of data.

 \begin{table*}[htbp!]
\caption{The SM prediction and experimental values of mass differences $\Delta m_F$ for $\Delta F=2$ mixing.
}
\begin{tabular}{ccc}\hline\hline
Mixing modes  &SM prediction   &Experimental data  \\\hline
$K-\bar K$  &  $4.7\pm1.8$ (ns)$^{-1}$~\cite{Brod:2011ty}&$5.293\pm0.009$(ns)$^{-1}$(PDG~\cite{PDG2024})
 \\\hline
$B_d- \bar B_d$ & $0.547^{+0.035}_{-0.046}$(ps)$^{-1}$~\cite{King:2019lal}& $0.5065\pm 0.0019$(ps)$^{-1}$(PDG~\cite{PDG2024})  \\\hline
$B_s- \bar B_s$ &  $18.23\pm0.63$(ps)$^{-1}$~\cite{King:2019lal,Albrecht:2024oyn}
&$17.765\pm 0.006$(ps)$^{-1}$(PDG~\cite{PDG2024}) \\\hline
\end{tabular}
\label{tab:mixing}
\end{table*}

The mass differences for the neutral meson
mixing system  are 
\begin{eqnarray}
    \Delta m_P=|\Delta m_P^{SM} - f_P^2 m_P \sum_i\tilde c_i(m_H)\eta_i(m_H)B_P^i(m_H)|.\nonumber\\
\end{eqnarray}
Therefore, the neutral meson mixing can provide the constraints for  the ALP parameter regions

\subsection{ ALP-Z boson  interaction}

For ALP couplings with the electroweak gauge bosons, the relevant interaction   can also be probed through precision measurements of the properties of Z bosons.

 Firstly, we focus on the exotic Z-boson decay $Z\to \gamma a$ induced by Eq.~(\ref{agauge}) at  tree level.
The decay rate can be obtained as
\begin{eqnarray}
    \Gamma(Z\to \gamma a)=\frac{m_Z^3}{384\pi}g_{a\gamma Z}^2\left(1-\frac{m_a^2}{m_Z^2}\right)^3\;.
\end{eqnarray}
Divided  the Z boson total decay width $\Gamma_Z=2.4955$GeV~\cite{PDG2024}, we obtain the branching ratios as 
\begin{eqnarray}
    Br(Z\to \gamma a)=2\times 10^{-3}\left(\frac{g_{a\gamma Z}}{2.8\times 10^{-3}}\right)^2\left(1-\frac{m_a^2}{m_Z^2}\right)^3\;.\nonumber\\
\end{eqnarray}
At 95\% CL, the  decay can be constrained by Z total decay width, which can be converted into $Br(Z\to inv)<2\times 10^{-3}$, which can constrain the corresponding model parameters, $m_a$ and $g_{aW}$.  
The most stringent constraints arise from the L3 search in $e^+e^-$ collisions at the Z resonance at LEP, where 
Br$(Z\to \gamma a)<1.1\times 10^{-6}$~\cite{L3:1997exg}, 
for photon energies exceeding 31 GeV. 

Furthermore, ALP can have the subsequent decays $a\to \gamma\gamma, l^+l^-$, which produces the  decay chains~\cite{OPAL:1991acn,ATLAS:2015rsn}
\begin{eqnarray}
	&& Br(Z\to \gamma ee)<5.2\times 10^{-4}\; (OPAL),
	\nonumber\\
	&& Br(Z\to \gamma \mu\mu)<5.6\times 10^{-4}\; (OPAL),
	\nonumber\\
	&& Br(Z\to \gamma \tau\tau)<7.3\times 10^{-4}\; (OPAL),
	\nonumber\\
	&& Br(Z\to \gamma \gamma\gamma)<2.2\times 10^{-6}\; (ATLAS).
\end{eqnarray}
These processes can place the constraints for the model parameters if kinematically allowed.

Additionally,  the ALP couplings can affect  electroweak precision observables at the loop level.
The loop corrections  can in general  be described in terms of the usual oblique parameters S,T,U~\cite{Peskin:1991sw}, with the forms as~\cite{Aiko:2023trb}
\begin{eqnarray}
S&=&\frac{c_W^2s_W^2}{72\pi^2m_Z^2}g_{aW}g_{aB}\left[F(m_Z^2;a,\gamma)-F(m_Z^2;a,Z)\right],\;\\
U&=&\frac{s_W^4g_{aW}^2}{72\pi^2m_Z^2}\nonumber\\
&\times &\left[F(m_Z^2;a,\gamma)+\frac{c_W^2}{s_W^2}F(m_Z^2;a,Z)-\frac{1}{s_W^2c_W^2}F(m_W^2;a,W)\right]\nonumber.
\end{eqnarray}
with the function $ F(k^2;a,V)$ defined as 
\begin{eqnarray}
&&3k^4\lambda(m_a/k,m_V/k)
 \times\left[B_0(k^2;m_a,m_V)-B_0(0;m_a,m_V)\right]\nonumber\\
 &&-3k^2\left[(2m_a^2+2m_V^2-k^2)B_0(0;m_a,m_V)\right]\nonumber\\
&&-3k^2\left[A_0(m_a)+A_0(m_V)\right]+7k^2(3m_a^2+3m_V^2-k^2)\;,
\end{eqnarray}
here the $A_0(m_0),B_0(p^2;m_0,m_1)$ means  Passarino-Veltman functions denoted explicitly as
\begin{eqnarray}
&&A_0(m_0)=m_0^2\left(1-\ln\frac{m_0^2}{\Lambda^2}\right),\\
&&B_0(p^2;m_0,m_1)=\int_0^1\mathrm{d}x\nonumber\\
&&\qquad \qquad \qquad \times\ln\left[\frac{\Lambda^2}{xm_0^2+(1-x)m_1^2-x(1-x)p^2}\right]\nonumber\;.
\end{eqnarray}

The current global fits for the oblique parameters are $S=-0.04\pm0.10$, $T=0.01\pm0.12$ and $U=-0.01\pm0.09$~\cite{PDG2024}. We found that within 1$\sigma$ errors, these parameters approaches zero.
The new-physics scale $\Lambda$ denotes the axion breaking scale.
By setting $\Lambda=10$TeV, S and U give the 
 the  bound correspondingly.

In views of the couplings $g_{a\gamma Z}$ at the tree level, we can consider the production of a photon in association with an ALP in  colliders. 
For the $e^+e^-$ colliders, the production process  proceeds via Z propogator in the s-channel with  the differential cross section as
\begin{eqnarray}
  \frac{d \sigma(e^+e^-\to \gamma a)} {d\cos\theta}&=&\frac{1}{512\pi}\frac{\alpha^2(s)}{\alpha(m_Z)}s^2\left(1-\frac{m_a^2}{s}\right)^3(1+\cos^2\theta)\nonumber\\
  &&\times[|V(s)|^2+|A(s)|^2]\;,
\end{eqnarray}
where $\sqrt{s}$ is the center-of-mass energy and $\theta$ denotes the scattering angle of the photon relative to the beam axis. Here we neglect the electron mass.
ALP emission from the initial-state leptons vanishes due to the loop-suppressed  $a-e-e$ interaction. The vector and axial-vector form factors are given by
\begin{eqnarray}
  &&  V(s)=\frac{1-4s_W^2}{4s_Wc_W}\frac{g_{a\gamma Z}}{s-m_Z^2+im_Z\Gamma_Z}+\frac{2g_{a\gamma\gamma}}{s},\nonumber\\
   &&A(s)=\frac{1}{4s_Wc_W}\frac{g_{a\gamma Z}}{s-m_Z^2+im_Z\Gamma_Z}\;.
\end{eqnarray}
Analyzing the vector coupling $V(s)$, the first term is suppressed by $1-4s_W^2$ so that we can reasonably ignore this term.  Additionally, it could have enhanced effect when $s\sim m_Z^2$.
Integrating out the angle $\theta$, we can obtain  the cross section
\begin{eqnarray}\label{enhance}
 \sigma(e^+e^-\to \gamma a)|_{s=m_Z^2}&=&\frac{\alpha(s)}{24}\left(1-\frac{m_a^2}{s}\right)^3 \nonumber\\
&\times & \left[\frac{m_Z^2}{\Gamma_Z^2}\frac{g_{a\gamma Z}^2}{64c_W^2 s_W^2}+(g^{eff}_{a\gamma\gamma})^2\right].
\end{eqnarray}
Note that the contribution in the case of Z pole receives an enhancement factor $m_{Z}^2/\Gamma_Z^2\approx 1330$.
It shows that using on-shell decays of narrow heavy SM particles into ALPs rather than the production of ALPs via an on-shell particle provides a much enhanced sensitivity to the $a\gamma Z$
 coupling on the Z pole.

For the  ALP in our case, Higgs boson can decays into Z bosons and ALPs,  $h\to Z a$ and $h\to a a$. However, the two decay processes are not induced by the Wilson coefficient $C_{WW}$ and  $C_{BB}$ so that we do not consider the relevant processes.

\section{ALP parameter bounds} \label{sec4}

Because the ALP couplings with fermions and gauge bosons depend on $g_{aW}$ and $g_{aB}$ at the same time, this means that both couplings will affect the phenomenology described above.
Therefore, we consider four different scenarios: the photophobic ALP,  ALP with $g_{aB}=0$, same sign ALP and ALP with $g_{aW}=0$. The first one turns off $g_{a\gamma\gamma}=0$ at tree-level by imposing the specific condition  $g_{aB}=-g_{aW}\tan^2\theta_W$ to eliminate ALP-photon interaction, while
 the third one  adopts the same sign $g_{aB}=g_{aW}\tan^2\theta_W$ to enhance ALP-photon interaction. The left two ones requires  $g_{aW,aB}=0$ directly.

In the following, we will analyze the current experimental bounds and future sensitivity for these four different scenarios, respectively.

\subsection{Photophobic ALP scenario $g_{a\gamma\gamma}=0$}

For the photophobic ALP scenario, the ALP coupling with the photon at the tree level is eliminated by the following condition:
\begin{eqnarray}\label{phot}
  &&  g_{aB}=-g_{aW}\tan^2\theta_W \longrightarrow g_{a\gamma\gamma}=0, \nonumber\\ 
 && g_{a\gamma Z}=2\tan\theta_W g_{aW}\;.
\end{eqnarray}

\begin{figure*}[!t]
\centering
 	\subfigure[\label{br_phot}]
 	{\includegraphics[width=.486\textwidth]{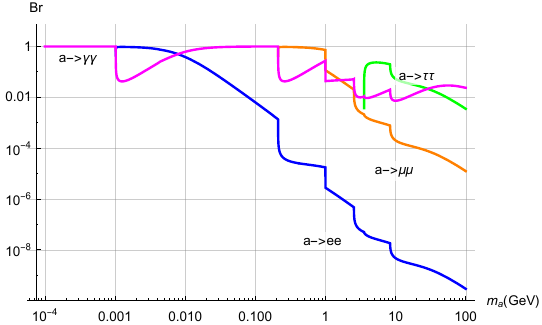}}
 	\subfigure[\label{br_hyper}]
 	{\includegraphics[width=.486\textwidth]{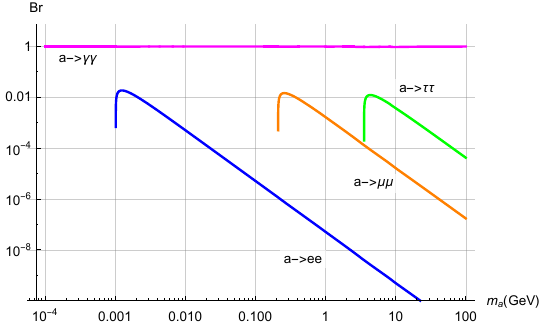}}
    \subfigure[\label{br_same}]
 	{\includegraphics[width=.486\textwidth]{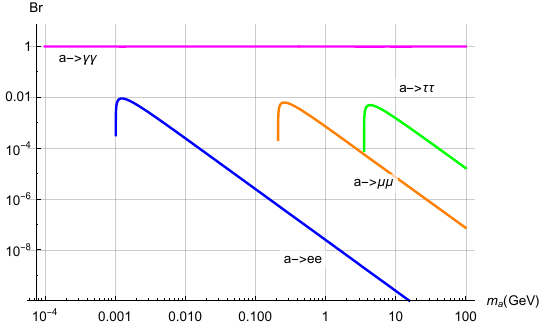}}
     \subfigure[\label{br_W}]
 	{\includegraphics[width=.486\textwidth]{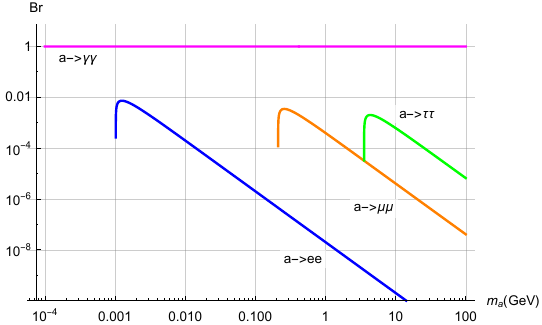}}
	\caption{ 
    The ALP branching ratios decays into different SM final states. The decays into photons, electrons, muons, and tauons are shown in magenta, blue, orange, and green, respectively.
    The upper left panel a) means the photobic ALP $g_{a\gamma\gamma}=0$. The upper right panel b) means the  ALP 
 with $g_{aB}=0$. The lower left panel  c) means the same sign case  $g_{aB}=g_{aW}\tan^2\theta_W$.
    The lower right panel  d) means the ALP  with $g_{aW}=0$.
	}
	\label{fig:Br}
\end{figure*}

In this case, the branching ratios for the corresponding decay chains can be determined, as shown in Fig.~\ref{br_phot}.
We found that below $m_a<2m_e$, the only decay channel is $a\to 2\gamma$. Additionally, the photon channels dominate the decay chains for $0.01<m_a<2m_\mu$ GeV and $m_a>25$ GeV again, even if they are doubly suppressed by  $m_a^3$ and $\alpha^2$.
When the ALP mass approaches the  double charged leptons, the dominant decay processes will convert into the corresponding leptons, with an order of magnitude proportional to the ALP mass $m_a$.
Furthermore, the branching ratios into charged leptons ($e,\mu$) decrease as the ALP mass increases, which significantly influences the shape of the parameter space contours.
These indicate that the ALP subsequent decays $M_1\to M_2 a(\to ll,2\gamma)$ can provide constraints on the model parameters, as summarized in Table~\ref{tab:decay}.

\begin{figure*}[!t]
\centering
 	\subfigure[\label{bound_phot}]
 	{\includegraphics[width=.486\textwidth]{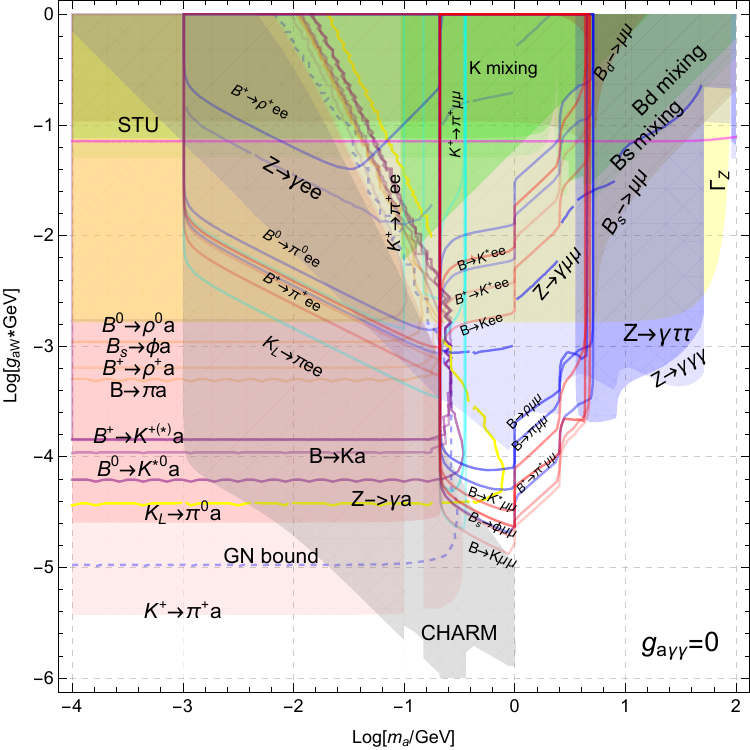}}
 	\subfigure[\label{bound_hyper}]
 	{\includegraphics[width=.486\textwidth]{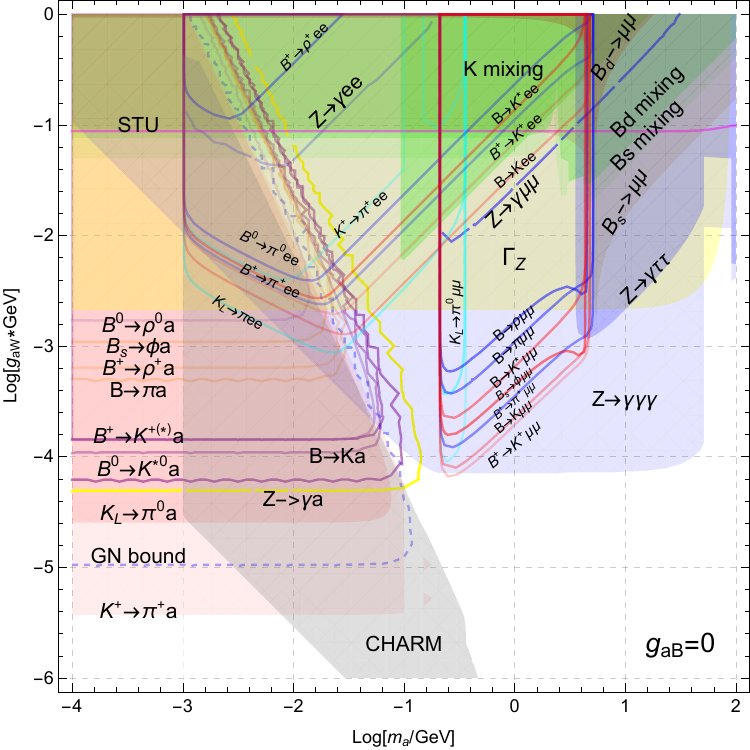}}
    \subfigure[\label{bound_same}]
 	{\includegraphics[width=.486\textwidth]{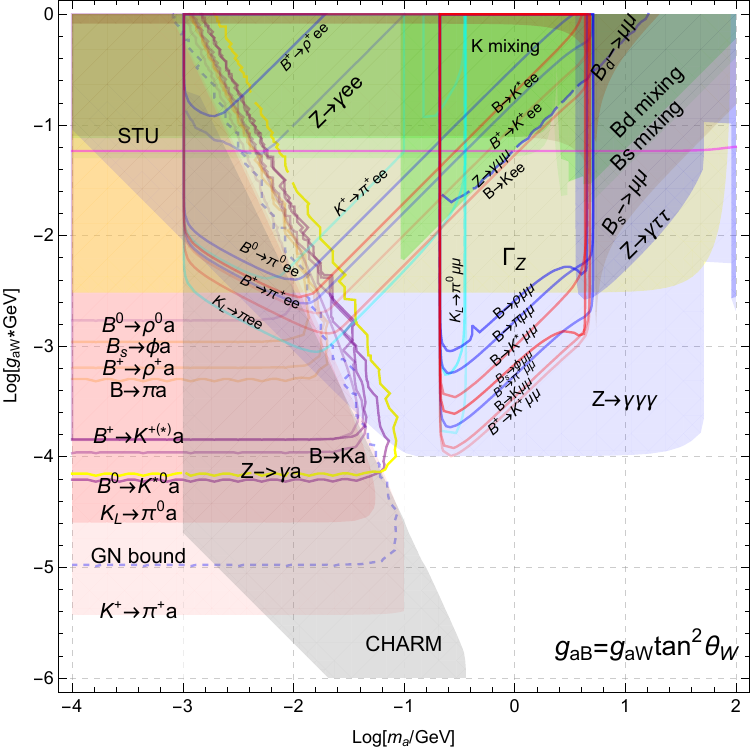}}
    \subfigure[\label{bound_W}]
 	{\includegraphics[width=.486\textwidth]{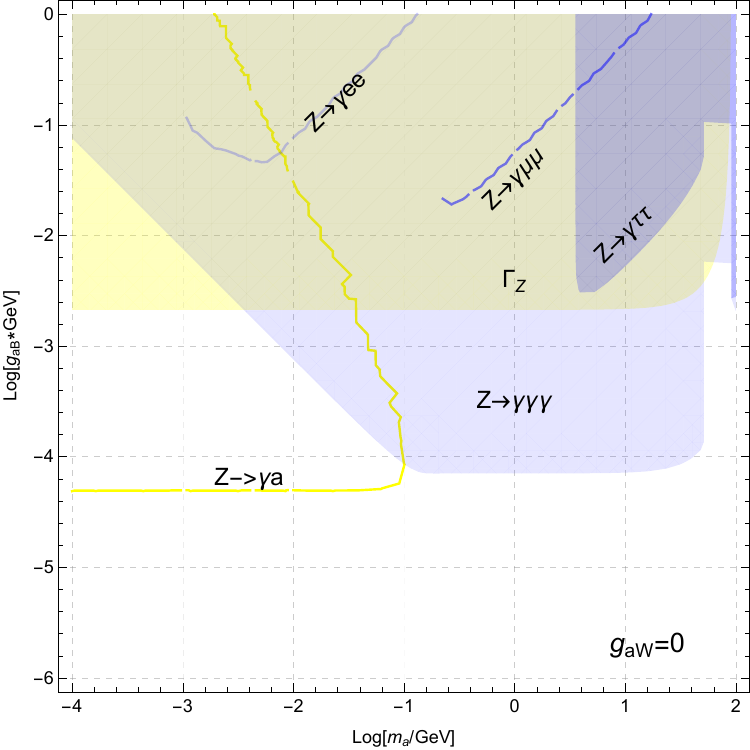}}
	\caption{ The excluded parameter regions from different physical processes in the plane $m_a-g_{aW}$. 
    The different physical processes exhibit distinctive exclusion capabilities, as shown in different colors.
   The upper left panel a) means the photobic ALP $g_{a\gamma\gamma}=0$. The upper right panel b) means the  ALP 
 with $g_{aB}=0$. The lower left panel  c) means the same sign case  $g_{aB}=g_{aW}\tan^2\theta_W$.
    The lower right panel  d) means the ALP  with $g_{aW}=0$.	}
	\label{fig:bound}
\end{figure*}

By inputting the photophobic form in Eq.~(\ref{phot}) into the above phenomenological physical processes and comparing it to the experimental observables in Table.~\ref{tab:decay}, we can obtain the corresponding exclusion parameter regions for $m_a$ and $g_{aW}$.
The parameter bounds are plotted in Fig.~\ref{bound_phot}. Here, we choose the ALP mass regions with  $10^{-4}<m_a<100$GeV.
The lower bound indicates $m_a<2m_e$, and the upper bound represents the electroweak scale.

We found that different physical processes exhibit distinctive exclusion abilities, as shown in different colors. For three different quark transitions, $s \to d$, $b \to d$, and $b \to s$, different decays demonstrate distinct exclusion capabilities.
Note that the physical processes $M_1 \to M_2 \nu\bar\nu$ should include the decay factor ($1 - \mathcal{P}_{\mathrm{dec}}^{a}$) to place the parameter constraints when ALP decays outside the detector.

For the $s \to d$ quark transition, the most stringent bound comes from $K^+ \to \pi^+ a$ with $g_{aW} > 10^{-5.4}$ GeV$^{-1}$ in the mass region $m_a < m_{K^+} - m_{\pi^+}$, which is stronger than $K_L \to \pi^0 a$ by an order of magnitude.
It indicates that the NA62 experiment provides a comparable exclusive limit to the previous E949 experiment.
Note that the gap around $m_a \in (100, 150)$ MeV is due to the pion mass pole from the NA62 Collaboration~\cite{NA62:2021zjw}.
Additionally, the GN bound provides stronger constraints, especially in the mass gap region and for large $g_{aW}$ shown in the blue dashed line.
It approaches $g_{aW} \sim 10^{-5}$ GeV$^{-1}$, providing much stronger constraints than $K_L \to \pi a$, but weaker constraints than $K^+ \to \pi^+ a$.

For $b \to d$ quark transition, $B \to \pi a$ provides the most stringent bounds with $g_{aW} < 10^{-3.3}$ GeV$^{-1}$ for $m_a < 0.16$ GeV. 
Correspondingly, the bound weakens as follows: $B^+ \to \rho^+ a$ with $g_{aW} < 10^{-3.2}$ GeV$^{-1}$, $B_s \to \phi a$ with $g_{aW} < 10^{-3}$ GeV$^{-1}$, and $B^0 \to \rho a$ with $g_{aW} < 10^{-2.8}$ GeV$^{-1}$. Additionally, large couplings with $g_{aW} > 10^{-2.6}$ results in a loss of distinguishable capability for $b \to d$ processes.

For $b \to s$ quark transition, $B^0 \to K^{0*} a$ provides the most stringent bounds with $g_{aW} < 10^{-4.2}$ GeV$^{-1}$ for $m_a < 4.5$ GeV. The limiting capability decreases sequentially by a factor of 1.6 from $B^0 \to K^{0*} a$, $B \to K a$, to $B^+ \to K^{+*} a$.

If $m_a > 2m_l$, the semi-leptonic decay $M_1 \to M_2 ll$ can occur kinematically. Therefore, the relevant decay processes can provide bounds on model parameters within reasonable regions in Fig.~\ref{bound_phot}. For $M_1 \to M_2 ee$, the strongest bounds come from $K_L \to \pi ee$ within the oval for $0.001 < m_a < 0.1$ GeV. 
The following bounds are from $B \to K ee$, $B^+ \to \pi^+ ee$, $B^+ \to K^+ ee$, $K^+ \to \pi^+ ee$, $B^0 \to \pi^0$, and $B^+ \to \rho^+$ within their respective circles.
For $M_1 \to M_2 \mu\mu$, $B^+ \to K^+ \mu\mu$ excludes $g_{aW}>10^{-4.8}$ GeV$^{-1}$, followed by
$B\to K\mu\mu$,
$B^+ \to \pi^+ \mu\mu$,  $B_s \to \phi \mu\mu$, $B\to K^*\mu\mu$, $B\to \pi\mu\mu$ and  $B\to \rho\mu\mu$.
Additionally, the upward-right tilt of the contour for  $M_1\to M_2 ll$ arises from the reduction in  $Br(a\to ll)$, implying that larger values of the coupling $g_{aW}$ are excluded by Eq.~(\ref{alepton}).
These bounds effectively complement the unexplored regions for $K^+ \to \pi^+ a$, particularly in the range $2m_l < m_a < m_{M_1}-m_{M_2}$.

Moreover, the purely leptonic decays $B_{d/s} \to ll$ can provide bounds shown in brown. Currently, experimental data indicate that muon final states impose stronger constraints than tauon cases. Analyzing $B_{d/s} \to \mu\mu$, we find that $B_d \to \mu\mu$ provides significantly weaker bounds than $B_s \to \mu\mu$, around $10^{1.6}$ orders of magnitude, which constrains $g_{aW} < 10^{-1}$ GeV$^{-1}$ at most mass ranges except $m_a\sim m_{B_s}$. Additionally, the excluded regions by $B_s \to \mu\mu$ fully encompass those by $B_d \to \mu\mu$.

Similarly, neutral meson mixings provide bounds shown in green. $B_s-\bar{B}_s$ mixing can place constraints across the ALP mass region with $g_{aW} < 10^{-1.4}$ GeV$^{-1}$. 
$B_d-\bar{B}_d$ mixing offers comparably weaker constraints. $K-\bar{K}$ mixing can achieve $g_{aW} \sim 10^{-2}$, which provides stronger bounds than $B_s-\bar{B}_s$ mixing, especially for $0.1 < m_a < 1$ GeV. Additionally, while neutral meson mixings provide weaker constraints compared to rare meson decays, they exclude some unexplored regions for meson decays.

For the Z boson properties, the bounds from Z boson  decay chains are shown in yellow. 
$\Gamma_Z$ excludes $g_{aW} > 10^{-2.8}$ GeV$^{-1}$ across the ALP mass region. 
$Z \to a\gamma$ can reach $g_{aW} \sim 10^{-4.2}$ and fills a small region between the yellow line and rare meson decays. 
And the excluded region is fully covered by $M_1\to M_2ll$ for $m_a>2m_\mu$.
Similarly, ALP can decay into $\gamma\gamma, ll$ to produce $Z \to 3\gamma, \gamma ll$. The relevant bounds are shown in blue. 
$Z \to \gamma ee$, $Z \to \gamma \mu\mu$, and $Z \to \gamma \tau\tau$ exclude regions within their respective capabilities. 
And $Z \to 3\gamma$ provides the most stringent bounds, encompassing all $Z \to \gamma ee, \mu\mu$ regions and a significant portion of $Z \to \gamma\tau\tau$.

Proton beam dump experiments searching for long-lived particles provide constraints complementary to those from direct searches for rare meson decays. 
Production occurs through rare decays $K \to \pi a$ and $B \to \pi/K a$, followed by the displaced decay $a \to \gamma\gamma,ee,\mu\mu$ within the detector. 
The strongest bound is from the CHARM experiment~\cite{CHARM:1985anb}, with the number of signals from ALP decays estimated in~\cite{Clarke:2013aya,Choi:2017gpf}.
\begin{eqnarray}
&&N_{d} \approx 2.9\times10^{17}\sigma
\cdot\mathrm{Br}\left(a\to \gamma\gamma,ee,\mu\mu\right)\nonumber\\
&&\qquad\times\left[\exp\left(-\Gamma_{a}\frac{480\mathrm{m}}{\gamma}\right)-\exp\left(-\Gamma_{a}\frac{515\mathrm{m}}{\gamma}\right)\right],\;\nonumber\\
&&\sigma=\frac{3}{14}Br\left(K^+\to\pi^++a\right)+\frac{3}{28}Br\left(K_L\to\pi^0+a\right)\nonumber\\
&&\quad +9\cdot10^{-8}Br\left(B\to X+a\right)\;.
\end{eqnarray}
Here $\gamma=10\mbox{GeV}/m_a$.  Since there is no signal from CHARM experiment, one then
finds the 90\% CL, $N_{d}<2.3$~\cite{Clarke:2013aya}.
The corresponding region is shown in gray in Fig.~\ref{bound_phot}. We observe that the CHARM experiment sets very strong bounds on the coupling $g_{aW}$. The CHARM shape exhibits three distinctive kick points, corresponding to the thresholds $2m_e$, $2m_\mu$ and  $3\pi$, respectively.  

\begin{figure*}[!t]
\centering
 	\subfigure[\label{photo_proj}]
 	{\includegraphics[width=.486\textwidth]{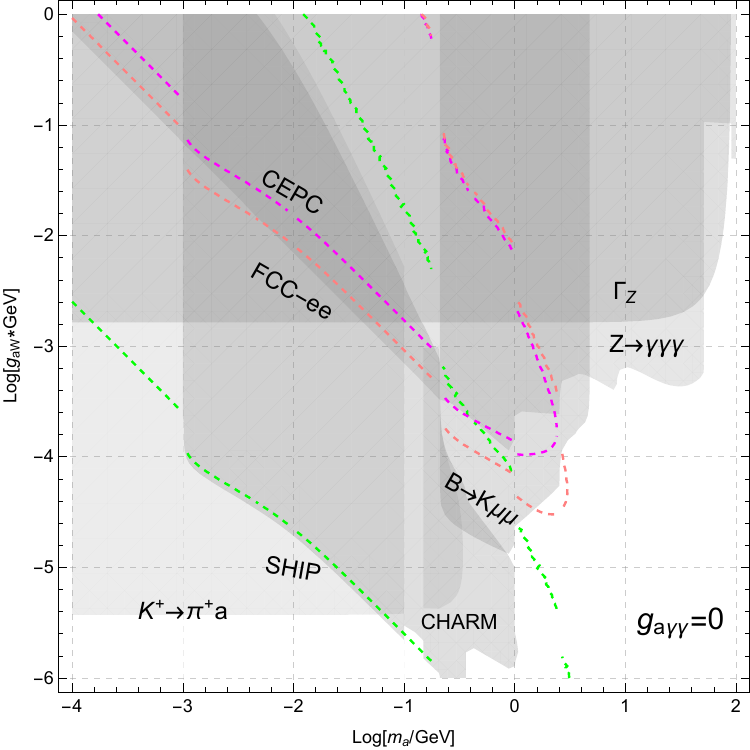}}
 	\subfigure[\label{hyper_proj}]
 	{\includegraphics[width=.486\textwidth]{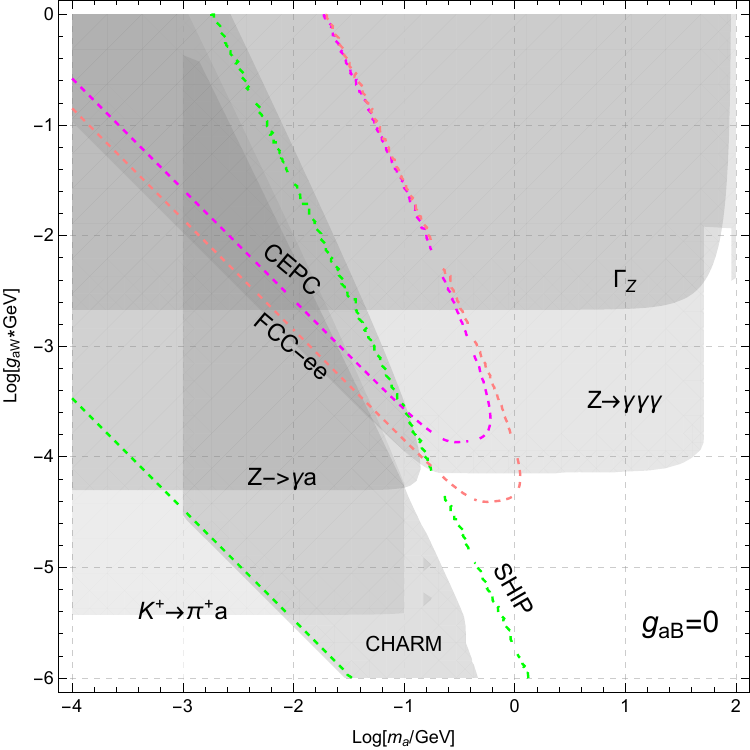}}
    \subfigure[\label{same_proj}]
 	{\includegraphics[width=.486\textwidth]{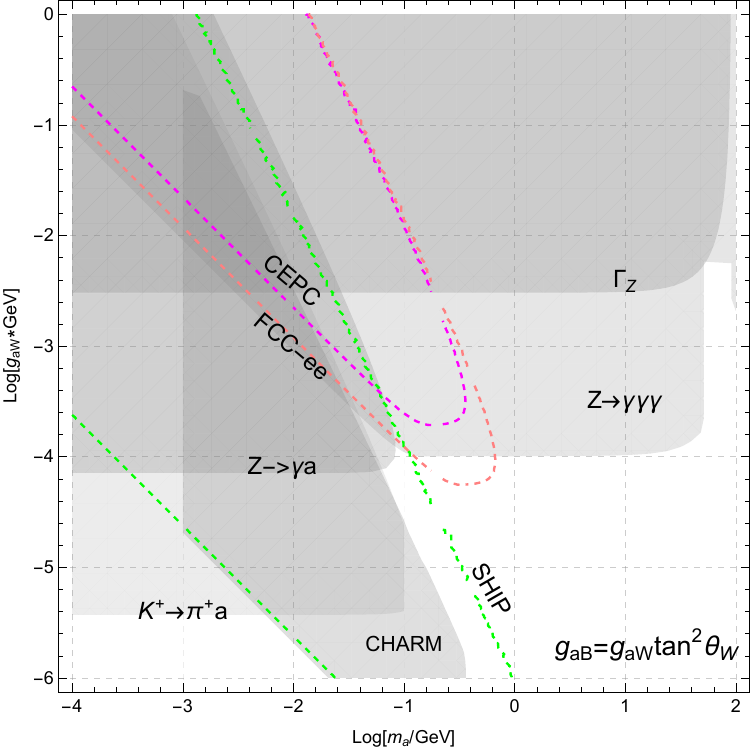}}
    \subfigure[\label{w_proj}]
 	{\includegraphics[width=.486\textwidth]{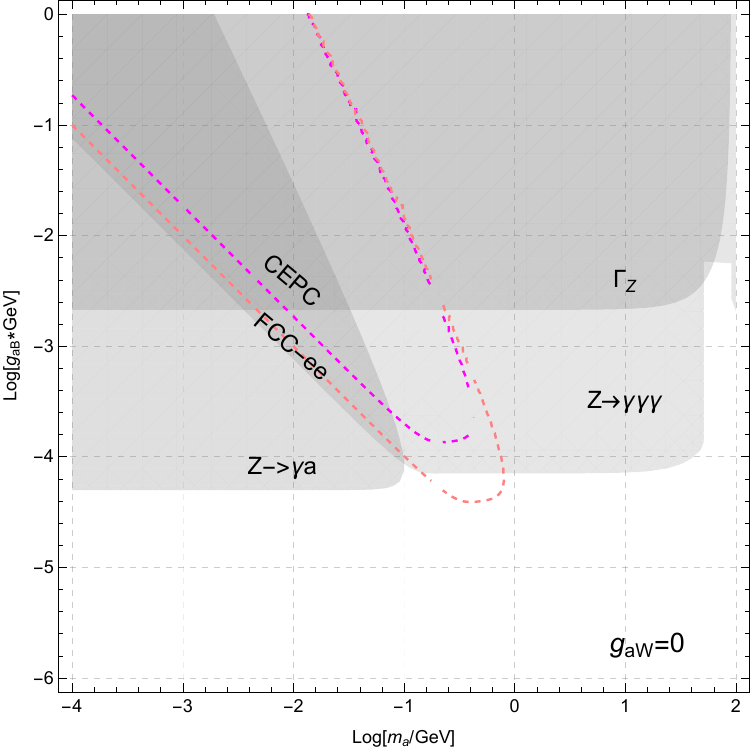}}
	\caption{ The future sensitivity for ALP parameter in the plane $m_a-g_{aW}$. 
    The future sensitivity of lepton colliders (CEPC,FCC-ee) and SHIP is shown in the dashed lines with magneta, orange and green, respectively.
 The upper left panel a) means the photobic ALP $g_{a\gamma\gamma}=0$. The upper right panel b) means the  ALP 
 with $g_{aB}=0$. The lower left panel  c) means the same sign case  $g_{aB}=g_{aW}\tan^2\theta_W$.
    The lower right panel  d) means the ALP  with $g_{aW}=0$.	}
	\label{fig:proj}
\end{figure*}

Therefore, we obtain the corresponding strongest bounds within the respective capability regions. These bounds are shown in gray in Fig.~\ref{photo_proj}, respectively.
In the following we want to analyze the future sensitivity for ALP parameter regions.

\begin{figure*}[!t]
\centering
 	\subfigure[\label{cross_phot}]
 	{\includegraphics[width=.486\textwidth]{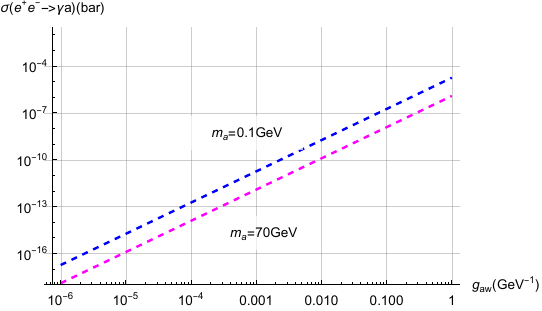}}
 	\subfigure[\label{cross_hyper}]
 	{\includegraphics[width=.486\textwidth]{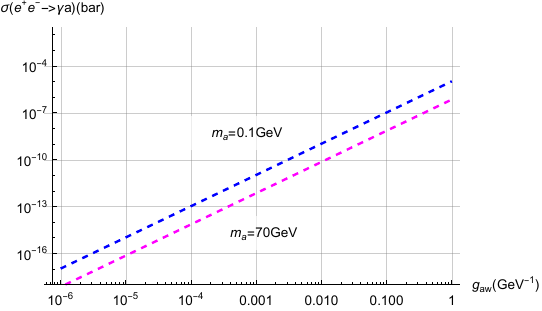}}
    \subfigure[\label{cross_same}]
 	{\includegraphics[width=.486\textwidth]{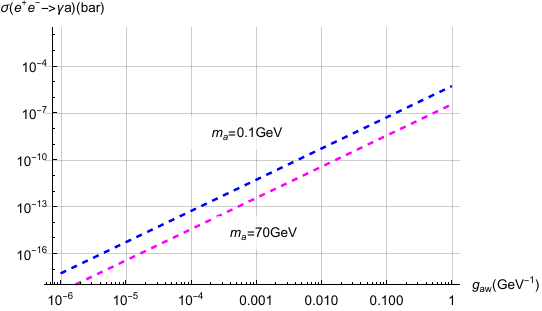}}
    \subfigure[\label{cross_W}]
 	{\includegraphics[width=.486\textwidth]{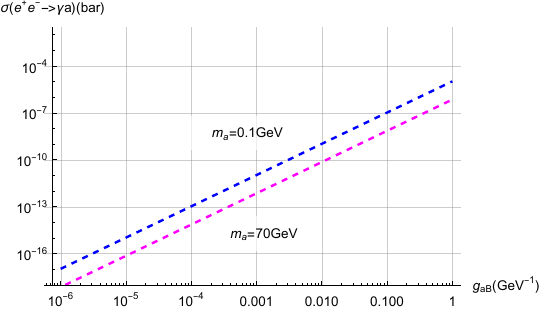}}
	\caption{ The cross section of $e^+e^-\to \gamma a$ with the coupling $g_{aW}$(GeV$^{-1}$) in the center of mass energy $\sqrt{s}=m_Z$. Here we illustrate two ALP mass choice, $m_a=0.5$GeV and $m_a=70$GeV.
   The upper left panel a) means the photobic ALP $g_{a\gamma\gamma}=0$. The upper right panel b) means the  ALP 
 with $g_{aB}=0$. The lower left panel  c) means the same sign case  $g_{aB}=g_{aW}\tan^2\theta_W$.
    The lower right panel  d) means the ALP  with $g_{aW}=0$.
	}
	\label{fig:bound_phot}
\end{figure*}

Firstly, we analyze the sensitivity of future lepton colliders for our model parameters. As indicated in Eq.~(\ref{enhance}), there is an enhancement factor $m_Z^2/\Gamma_Z^2$ particularly relevant for future colliders operating at the Z pole energy $\sqrt{s}=m_Z$. This highlights the importance of future lepton collider analyses at $\sqrt{s}=m_Z$.
Fortunately, based on the conceptual design, two $e^+e^-$ colliders with $\sqrt{s}=91$ GeV exist: FCC-ee~\cite{FCC:2018evy} and CEPC~\cite{CEPCStudyGroup:2018rmc}. At this center-of-mass energy, the corresponding luminosities are 192 ab$^{-1}$  and  16 ab$^{-1}$, respectively.
 This implies that the sensitivity of future lepton colliders can be employed to constrain the parameter space. 
 The corresponding cross section is shown in Fig.~\ref{cross_phot}. We found that for $g_{aW}\sim 10^{-2}$, the cross section approaches $\sigma(e^+e^-\to \gamma a)\sim(10^{-10}-10^{-8})$ barn, depending on the ALP mass.

By further considering the decay of ALPs within the detector, we calculate the number of signal events as
 \begin{eqnarray}
 && N_{sig}=\sigma(e^+e^-\to \gamma a)\cdot \ell_{lum}
  (\mathcal{P}_{\mathrm{dec}}^{a}(l_D)-\mathcal{P}_{\mathrm{dec}}^{a}(r)).
 \end{eqnarray}
Here, $\ell_{lum}$ represents the luminosity of the lepton collider, 
and $r(l_D)$ denotes the minimal and maximal distances from the interaction point (IP) at which the detector can detect an ALP decay into SM particles.
The main detector sensitivity is chosen as $r=5$ mm and $l_D=1.22$ mm~\cite{Chrzaszcz:2020emg}, respectively. 
Requiring $N_{sig} \geq 3$, we obtain the future sensitivity shown as a dashed line in Fig.~\ref{photo_proj}, for CEPC in magenta and FCC-ee in orange, respectively. 
We found that FCC-ee provides better sensitivity than CEPC, which is approximately close to the $Z \to \gamma\gamma\gamma$ parameter regions. Furthermore, FCC-ee can even reach $g_{aW} \sim 10^{-4.5}$ GeV$^{-1}$ for $m_a \sim 10^{0.4}$ GeV.
Furthermore, the two colliders can cross check the bounds from  $B\to K\mu\mu$.
Additionally, the FCC-ee and CEPC lose their discriminative capabilities for   $m_a>10^{-0.8}$ GeV and $g_{aW}>10^{-4}$GeV$^{-1}$.

The Search for Hidden Particles (SHiP)~\cite{SHiP:2021nfo} is an approved beam-dump experiment scheduled to begin operation in 2031. At SHiP, a 400 GeV proton beam extracted from the CERN SPS accelerator impacts a heavy proton target, resulting in significant production rates of pseudoscalar mesons $K, B, B_s$. These produced mesons can be utilized to search for ALPs through rare meson decays.
SHIP can produce the total meson numbers $8.1\times10^{13}$ for $B^{0,\pm}$ and $2.16\times10^{13}$ for $B_s$~\cite{Bondarenko:2018ptm,Wang:2024mrc}, respectively.
Assuming these mesons decay into ALPs within the detector, leading to observable signals, we can obtain~\cite{Chrzaszcz:2020emg}
\begin{eqnarray}
    N_{sig}&=&(N_{B}\sum_i (Br(B\to M_i a)
    +N_{B_s} (Br(B_s\to \phi a))\nonumber\\
    &\times & \left[\exp\left(-\Gamma_a\frac{l}{\gamma}\right)-\exp\left(-\Gamma_a\frac{l+\Delta l}{\gamma}\right)\right]\;,
\end{eqnarray}
here $\gamma=25\text{GeV}/m_a$, $i=\pi, K, K^*, \rho$ and their corresponding charged components are considered. The detector size is $\Delta l=55$ meters and  $l= 70$ m. Similarly, requiring $N_{sig} \geq 3$, we obtain the future projection shown as a green dashed line in Fig.~\ref{photo_proj}. We find that SHiP provides significant sensitivity, particularly for small $g_{aW}$. 
Furthermore, the SHiP sensitivity projection completely encompasses the CHARM exclusion limits.
Therefore, SHiP and FCC-ee serve as complementary explorations for ALPs, focusing on distinct parameter regions.

Additionally, the Drell-Yan process $pp \to \gamma a$ at the LHC can provide bounds for $m_a > 100$ GeV, as mentioned in Ref.~\cite{Craig:2018kne}, though this is beyond the mass range of interest in our study.

\subsection{ ALP scenario  with $g_{aB}=0$}

Another interesting scenario is the  ALP with $g_{aB}=0$, which couples only with $SU(2)_L$ gauge bosons as
\begin{eqnarray}\label{hyper}
    g_{aB}=0 \longrightarrow g_{a\gamma\gamma}=g_{aW}s_W^2,\; 
    g_{a\gamma Z}=2c_W s_W g_{aW}.
\end{eqnarray}
Correspondingly, ALP interaction with fermions at loop level is obtained as 
\begin{eqnarray}
g_{aFF}&=&\frac{9\alpha g_{aW}}{64\pi s_W^2}\log\frac{\Lambda^2}{m_W^2}+\frac{3}{2}Q_F^2 \frac{\alpha}{4\pi}g_{a\gamma\gamma}\log\frac{m_W^2}{m_F^2}\;.
\end{eqnarray}
Note that the tree-level $g_{a\gamma\gamma}$ is significantly larger than the loop contribution $g^{loop}_{a\gamma\gamma}$ and $g_{aFF}$, by approximately one order of magnitude. 
The associated branching ratios for ALP decay processes are  drawn in Fig.~\ref{br_hyper}. 
We find that regardless of the ALP mass, the dominant decay channel is $a \to \gamma\gamma$ with a branching ratio close to 1. 
Additionally, the chains decaying into charged leptons are strongly suppressed even if kinematically allowed. This feature differs significantly from the photophobic case. This suggests that different parameter bounds could exist in the  ALP scenario with $g_{aB}=0$.

Similarly, using Eq.~(\ref{hyper}) to analyze physical processes and comparing them with the experimental observables in Table.~\ref{tab:decay}, the corresponding exclusion parameter regions for $g_{aW}$ are shown in Fig.~\ref{bound_hyper} for $10^{-4} < m_a < 100$ GeV.

The different physical processes show distinct exclusion abilities represented by different colors. We find that the largest upper bound for $g_{aW}$ maintains the same constraints as the photophobic ALP scenario. However, the corresponding excluded regions are narrowed due to the enhanced total decay width, affecting the decay factor ($1-\mathcal{P}_{\mathrm{dec}}^{a}$).

For rare meson two-body decays $M_1 \to M_2 a$, three different quark transitions—$s \to d$, $b \to d$, and $b \to s$—will move left as the coupling $g_{aW}$ increases. 
For instance, when $g_{aW} = 1$ GeV$^{-1}$, the excluded mass changes from $10^{-2.2}$ GeV in the photophobic scenario to $10^{-3}$ GeV.
For the $s \to d$ quark transition, the lowest excluded value remains consistent with the photophobic case at $g_{aW} \sim 10^{-5.4}$ GeV$^{-1}$.
Note that the original gap around the pion mass disappears approximately because the weakened exclusion capability constrains $m_a < 10^{-1}$ GeV.
The excluded regions by the GN bound lie between $K_L \to \pi^0 a$ and $K^+ \to \pi^+ a$. Both reduce correspondingly to $m_a < 10^{-1}$ GeV.
Similarly, the constrained ALP mass decreases from $10^{-0.6}$ GeV to $10^{-1}$ GeV for $b \to s$ quark transition, and from $10^{-0.8}$ GeV to $10^{-1.6}$ GeV for $b \to d$ quark transition.
Correspondingly, the bound ability weakens similarly to the photophobic scenario, followed by $B \to \pi a$, $B^+ \to \rho^+ a$, $B_s \to \phi a$, and $B^0 \to \rho a$ for $b \to d$ transition. 
The limiting capability decreases sequentially from $B^0 \to K^{0*} a$, $B \to K a$ to $B^+ \to K^{+(*)} a$ for $b \to s$ transition. Additionally, the exclusion capability becomes indistinguishable for large coupling $g_{aW}$.

For the semi-leptonic decay $M_1 \to M_2 ll$, the constraints weaken when kinematically allowed decays occur. The bounded regions are illustrated in Fig.~\ref{bound_hyper}.
For the electron case, the bounded circle shifts to the upper left panel, corresponding to $g_{aW} \sim 10^{-3}$, as given by $K_L \to \pi ee$.
The subsequent bounds are derived from $B \to K ee$, 
$B \to K^* ee$, $B^+ \to K^+ ee$, 
$B^+ \to  \pi^+ ee$,  $B \to K* ee$, $K^+ \to \pi^+ ee$, $B^0 \to \pi^0$, and $B^+ \to \rho^+$, each showing varying degrees of weakening.
For muon cases, the corresponding bounds  shift upward, indicating that $M_1 \to M_2 ll$  impose weaker  bounds on the model parameters, reduced by an order of magnitude of  $10^{-0.6}$  compared to   the photophobic case.

For the case of pure leptonic decays $B_{d/s} \to ll$, the bounds remain approximately the same in brown with the photophobic case due to minor modifications in $g_{aFF}$. 
$B_s \to \mu\mu$ excluded region fully encompasses the ones by $B_d \to \mu\mu$ with a decreased $g_{aW}$. Additionally, neutral meson mixings remain unchanged as the process solely depends on the coupling $g_{ad_i d_j}$ induced by $g_{aW}$.

For the Z boson properties, the bounds from Z boson invisible decays increase slightly due to the modification $g_{a\gamma Z}$ from $2\tan\theta_W g_{aW}$ to $2c_W s_W g_{aW}$ as shown in yellow. Additionally, $Z \to a\gamma$ restricted interval decreases from $m_a \sim 10^{-0.2}$ GeV to $m_a \sim 10^{-0.8}$ GeV, indicating that the excluded regions do not intersect with $M_1\to M_2 \mu\mu$.
This feature is obviously different from the photophobic scenario.
This helps explore small regions in previously unexplored  meson decays.
Additionally, the exclusion capability of $Z \to \gamma ll$ weakens to varying degrees. For instance, $Z \to \gamma\tau\tau$ constrains $g_{aW} < 10^{-2.8}$ GeV$^{-1}$, which is weaker than the previous bound $g_{aW} < 10^{-3.4}$ GeV$^{-1}$ in the photophobic case. 
However, the enhanced decay ratio $a \to \gamma\gamma$ strengthens the corresponding bounds to around $g_{aW} < 10^{-4.2}$ GeV$^{-1}$, particularly for $m_a \in (10^{-1}, 10^{1.6})$ GeV.
Furthermore, $Z \to 3\gamma$ covers all regions excluded by $Z \to \gamma ll$.

Proton beam dump experiments (CHARM) have already excluded some regions in gray with a corresponding leftward shift. The excluded region by CHARM covers all coupling ranges $g_{aW} \in (10^{-6}, 1)$ GeV$^{-1}$. In the following, we analyze the future sensitivity for ALP parameter regions.

For the future lepton collider (CEPC, FCC-ee) operating at the Z pole with $\sqrt{s} = m_Z$, the corresponding cross section is shown in Fig.~\ref{cross_hyper}. We observe that the cross section only undergoes minor modifications. For example, $\sigma(e^+e^-\to \gamma a) \sim 10^{-7}$ bar for $m_a = 0.1$ GeV and $g_{aW} = 0.1$ GeV$^{-1}$.

After analyzing the ALP decay within the detector and requiring the signal events to be larger than 3, we obtain the future sensitivity shown as a dashed line in Fig.~\ref{hyper_proj}, for CEPC in magenta and FCC-ee in orange, respectively. 
We find that the CEPC sensitivity has already fully overlapped with $Z\to 3\gamma$. 
FCC-ee can provide significantly better sensitivity, touching the coupling range $g_{aW} \in (10^{-4.4}, 10^{-4.2})$ GeV$^{-1}$ for $10^{-0.8} < m_a < 10^{0.1}$ GeV. 
On the other hand, SHiP can produce a large number of pseudoscalar mesons decaying into ALP with sensitivity shown in green dashed line. 
This indicates that SHiP can provide strong sensitivity, especially for small $g_{aW}$. This shows that SHiP and FCC-ee can serve as complementary explorations for ALP, focusing on different parameter regions, respectively.

\subsection{Same sign ALP scenario $g_{aB}=g_{aW}\tan^2\theta_W$}

 Instead of selecting the opposite-sign coupling $g_{aB}=-g_{aW}\tan^2\theta_W$  to cancel out $g_{a\gamma\gamma}$ in the photophobic scenario,adopt the same-sign  coupling $g_{aB}=g_{aW}\tan^2\theta_W$ to enhance $g_{a\gamma\gamma}$. We  defined the case as same sign ALP scenario with 
\begin{eqnarray}\label{same}
  &&  g_{aB}=g_{aW}\tan^2\theta_W \longrightarrow g_{a\gamma\gamma}=2 g_{aW}s_W^2, \nonumber\\ 
 && g_{a\gamma Z}=2\tan\theta_W g_{aW}(1-2s_W^2)\;.
\end{eqnarray}

Correspondingly, ALP interaction with fermions at loop level is obtained as 
\begin{eqnarray}
g_{aFF}&=&\frac{3\alpha }{16\pi}g_{aW}\left[\frac{3}{4s^2_W}+\frac{(Y_{F_L}^2+Y_{F_R}^2)}{c^4_W}s_W^2\right]\log\frac{\Lambda^2}{m_W^2}\nonumber\\
&+&\frac{3}{2}Q_F^2 \frac{\alpha}{4\pi}g_{a\gamma\gamma}\log\frac{m_W^2}{m_F^2}\;.
\end{eqnarray}
Note that the tree-level $g_{a\gamma\gamma}$ is  four times larger compared the  ALP scenario with $g_{aB}=0$, and it is further significantly larger than both the loop contribution  $g^{loop}_{a\gamma\gamma}$ and $g_{aFF}$. 
The associated branching ratios for ALP decay processes are  drawn in Fig.~\ref{br_same},  indicating that the dominant decay channel is $Br(a \to \gamma\gamma)\sim 1$ across the entire ALP mass range.
Additionally, the decay into charged leptons is strongly suppressed, with ratios below 0.01. This feature is similar to the $g_{aB}=0$ scenario, which yields similar parameter bounds.

Based on Eq.~(\ref{same}) to analyze physical processes and compare with the experimental observables in Table.~\ref{tab:decay}, the corresponding exclusion parameter regions for $g_{aW}$ are shown in Fig.~\ref{bound_same} for $10^{-4} < m_a < 100$ GeV.

The different physical processes show distinct exclusion abilities represented by different colors. We find that the bounds from meson decays, including $M_1\to M_2 a$, $M_1\to M_2 ll$ and $M_1\to  ll$, keep similar constraints to those in the   ALP scenario with $g_{aB}=0$. 
This is due to the loop-suppressed ALP couplings with fermions and photons, which results in $a\to \gamma\gamma$ being the dominant decay channel from the tree-level $g_{a\gamma\gamma}$.

Additionally, the other bounds maintain comparable exclusion capabilities, such as meson mixing and CHARM.
The only significant changes involve Z boson-related processes. For the total decay width of the Z boson, the exclusion bound is modified to $g_{aW}<10^{-2.6}$, improved from $g_{aW}<10^{-2.8}$ in the $g_{aB}=0$ scenarios as shown in yellow.
Furthermore, $Z \to a\gamma$ restricted interval decreases from  $m_a \sim 10^{-0.8}$ GeV to $m_a \sim 10^{-1.2}$ GeV, which adds to previously unexplored rare meson decays.
Additionally, the exclusion capability of $Z \to \gamma ll$ weakens to varying degrees as shown in blue. For instance, $Z \to \gamma\tau\tau$ constrains $g_{aW} < 10^{-2.6}$ GeV$^{-1}$, which matches the exclusion capability of  $\Gamma_Z$.
Correspondingly, $a \to \gamma\gamma$ weakens the corresponding bounds to around $g_{aW} < 10^{-4}$ GeV$^{-1}$  for $m_a \in (10^{0.4}, 10^{1.6})$ GeV.
Furthermore, $Z \to 3\gamma$ covers all regions excluded by $Z \to \gamma ll$, even including the regions excluded by  $\Gamma_Z$. This feature differs slightly from the $g_{aB}=0$ scenario.

In the following, we analyze the future sensitivity for ALP parameter regions.
For the future lepton collider (CEPC, FCC-ee) operating at the Z pole with $\sqrt{s} = m_Z$, the corresponding cross section is shown in Fig.~\ref{cross_same}, keeping the same cross section approximately.

Similarly, by analyzing the ALP decay within the detector and requiring the signal events to be greater than 3, we obtain the future sensitivity shown as a dashed line in Fig.~\ref{same_proj}, for CEPC in magenta and FCC-ee in orange. We find that the sensitivity shifts to the left, and the FCC-ee sensitivity can fully cover the CEPC sensitivity.
Additionally, the $Z\to 3\gamma$ bound lies between the CEPC and FCC-ee sensitivities.
On the other hand, SHiP sensitivity  follows a similar trend of shifting to the left, providing strong sensitivity for small $m_a<1$ GeV. This shows that SHiP and FCC-ee can serve as complementary explorations for ALP, focusing on different parameter regions, respectively.

\subsection{ALP scenario  with $g_{aW}=0$}

Correspondingly, another left scenario is $g_{aW}=0$ ALP, which only couples with $U(1)_B$ hypercharge gauge boson as
\begin{eqnarray}\label{W}
  &&  g_{aW}=0 \longrightarrow g_{a\gamma\gamma}= g_{aB}c_W^2, \nonumber\\ 
 && g_{a\gamma Z}=-2\cos\theta_W\sin\theta_W g_{aB}\;.
\end{eqnarray}

Similarly, ALP interaction with fermions at loop level is obtained as 
\begin{eqnarray}
g_{aFF}&=&\frac{3\alpha }{16\pi}g_{aB}\left[\frac{(Y_{F_L}^2+Y_{F_R}^2)}{c^2_W}\right]\log\frac{\Lambda^2}{m_W^2}\nonumber\\
&+&\frac{3}{2}Q_F^2 \frac{\alpha}{4\pi}g_{a\gamma\gamma}\log\frac{m_W^2}{m_F^2}\;.
\end{eqnarray}
Note that the tree-level $g_{a\gamma\gamma}$ remains significantly larger than both the loop contribution  $g^{loop}_{a\gamma\gamma}$ and $g_{aFF}$. 
The associated branching ratios for ALP decay processes are  drawn in Fig.~\ref{br_W},  demonstrating  that the dominant decay channel is $Br(a \to \gamma\gamma)\sim 1$ across the entire ALP mass range.
Furthermore, the decay branching ratios into charged leptons are significantly suppressed to below 0.01. This feature is consistent with other scenarios, which yield similar parameter constraints.

Based on Eq.~(\ref{W}) to analyze physical processes and compare with the experimental observables in Table.~\ref{tab:decay}, the corresponding exclusion parameter regions for $g_{aW}$ are presented in Fig.~\ref{bound_same} for $10^{-4} < m_a < 100$ GeV.
Different physical processes exhibit distinct exclusion capabilities, represented by various colors.
Note that for $g_{aW}=0$, the  flavor changing interactions in Eq.~\ref{Eq:axion_interaction} vanish, causing  the relevant bounds ($M_1\to M_2 a$, $M_1\to M_2 ll$ and meson mixings) to disappear.
The only remaining physical processes are Z boson-related interactions.
For $m_a<0.1$GeV, the strongest bounds originate from $Br(Z\to \gamma a)$ with $g_{aB}<10^{-4.3}$GeV$^{-1}$.
For $m_a>0.1$GeV, the stringent constraint arises from $Z\to 3\gamma$ with $g_{aB}<10^{-4.2}$GeV$^{-1}$.
Other subsequent decays, such as $Z\to all$, provide relatively weak constraints, which are entirely covered by the total decay width of the Z boson, $\Gamma_Z$, as indicated in yellow.

For the future sensitivity of ALP parameter regions, the projected performance of future lepton colliders (e.g., CEPC and FCC-ee) operating at the Z pole ($\sqrt{s} = m_Z$) is shown in Fig.~\ref{cross_W}.
Similarly, by analyzing ALP decays within the detector and requiring at least three signal events, we derive the projected future sensitivity shown as dashed lines in Fig.~\ref{w_proj}, with CEPC in magenta and 
FCC-ee in orange.
We find that FCC-ee offers significantly better sensitivity compared to CEPC.
Additionally, they can act as complementary probes for ALPs, in comparison to $Z$ boson decays.

\section{conclusion} \label{sec5}

 We provide an updated and extended analysis on the experimental limits of the ALP couplings to electroweak gauge bosons across the ALP mass range from MeV to 100 GeV.
The ALP coupling with electroweak gauge bosons can generate the flavor-changing quark interaction exclusively through one-loop diagrams involving the W bosons within the framework of minimal flavor violation. 
Leveraging independent quark masses and CKM unitarity, the one-loop Feynman diagram with $W^\pm$ bosons propagating generates the quark-ALP coupling $g_{ad_id_j}$, which is ultraviolet finite for different flavor indices $i \neq j$. 
This new flavor-changing coupling, distinct from direct ALP-quark interactions, warrants further phenomenological investigation. 
Current experimental constraints from flavor-changing processes, such as rare meson decays and neutral meson mixing, provide significant bounds. 
The relevant experimental constraints are illustrated in Fig.~\ref{fig:bound}. 
We clearly depict the bounds from each different physical process. This helps us identify which process contributes to which specific constraint. With more precise future measurements of certain processes, we will be able to understand how the constraints on the parameter space evolve.

The flavor-changing down-quark interactions involve three distinct couplings corresponding to the quark transitions $s \to d$, $b \to d$, and $b \to s$, respectively.
Analysis of all relevant experimental constraints reveals that rare meson decays $M_1 \to M_2 a$ provide more stringent limits than meson mixing in kinematically allowed ALP regions with $m_a < m_{M_1} - m_{M_2}$.
For the three quark transition scenarios $s \to d$, $b \to d$, and $b \to s$, the strongest bounds are provided by $K^+ \to \pi^+ a$, $B \to \pi a$, and $B^0 \to K^{0*} a$, respectively.
Furthermore, the strongest bounds could reach a sensitivity of approximately $g_{aW} < 10^{-5.4}$ GeV$^{-1}$, as given by $K^+ \to \pi^+ a$.
And one-loop evolution can generate flavor conserving axion-fermion couplings, inducing subsequent decays $a \to ll$, with $M_1 \to M_2 ll$ providing constraints on the model parameters.
Correspondingly, the purely-leptonic decay can place bounds on model parameters.
Additionally, ALP couplings with the Z boson naturally emerge, motivating investigations into Z boson measurements, including the invisible decay $Z \to a\gamma$, $Z \to 3\gamma, \gamma ll$, and STU oblique parameters.
We find that $Z \to a\gamma$ can provide comparable constraints beyond the kinematically allowed mass range for rare meson decays.
The STU oblique parameters yield weaker constraints around $g_{aW} \sim 0.1$ GeV$^{-1}$ across the entire ALP mass region, which is independent of ALP mass basically.
The $Z \to 3\gamma$ decay can provide stringent constraints, especially for large ALP mass $m_a > 0.1$ GeV, which belongs to the unexplored regions for rare meson decays.
 It indicates that rare meson decays and $Z$ boson decays can serve as complementary explorations of ALP parameter regions, acting on the MeV and GeV mass scales of ALP, respectively.
 Furthermore, CHARM has excluded a large portion of parameter regions by searching for signal events.

The gauge-invariant ALP couplings to electroweak gauge bosons have  two independent interactions, namely the ALP-$SU(2)_L$ gauge boson coupling $g_{aW}$ and the ALP-hypercharge gauge boson coupling $g_{aB}$.
 In order to obtain the detailed parameter bounds, we analyze four different scenarios: photophobic $g_{a\gamma\gamma}=0$,  $g_{aB}=0$,  same sign ALP $g_{aB}=g_{aW}\tan^2\theta_W$ and  $g_{aW}=0$.
The first scenario involves suppress the tree level $g_{a\gamma\gamma} = 0$ by choosing specific couplings with $SU(2)_L$ and $U(1)_Y$ gauge bosons, effectively setting $g_{aB} = -g_{aW} \tan^2\theta_W$.
The second scenario requires that the coupling with $U(1)_Y$ hypercharge field vanishes directly, i.e., $g_{aB} = 0$. 
The third scenario focus on same sign coupling to enhance the tree level $g_{a\gamma\gamma}$.
The last scenario turns off the flavor-changing interactions directly. 
Correspondingly, the ALP parameter bounds for the four scenarios are shown in Fig.~\ref{fig:bound}.
The last three scenarios results in a larger ALP decay width due to the presence of the coupling with photon at the tree level.
This enhanced photon coupling naturally affects the relevant physical processes, resulting in the parameter bounds being significantly different between the four scenarios.
Comparing these four scenarios, we find that the excluded region in the photophobic case is significantly wider than in the other three cases.
The corresponding excluded regions shift to the left panel for these four scenarios due to the increased total ALP decay width.
Another distinct difference is that the bounds from $M_1 \to M_2 ll$ move upper as the scenario transitions from photophobic and $g_{aB}=0$ ALPs to same-sign ALPs, even vanish in the ALPs with $g_{aW}=0$,  indicating a gradual weakening of constraints.
Furthermore, $Z \to 3\gamma$ in the last three scenario provides much stricter bounds than in the photophobic ALP case, owing to the enhanced branching ratio $a \to \gamma\gamma$.

The future experimental projections for ALP parameters are analyzed as shown in Fig.~\ref{fig:proj}, including lepton colliders (CEPC, FCC-ee) and the Search for Hidden Particles (SHiP).
For lepton colliders operating near the Z pole, the ALP production via $e^+e^-\to Z\to \gamma a$ experiences an enhancement factor of $m_Z^2/\Gamma_Z^2=1330$, requiring an analysis of the corresponding projection.
We find that FCC-ee can provide better sensitivity than CEPC, allowing for probing regions beyond the bounds from rare meson decays and Z boson decays.
Additionally, lepton colliders can complement their sensitivity with the decay $Z\to 3\gamma$.
Moreover, SHiP can explore deeper into smaller $g_{aW}$ values, fully surpassing the existing CHARM bounds.
These indicate that future lepton colliders and SHiP can offer enhanced sensitivity for ALP parameters.

\section*{Acknowledgments}
We thank Prof. YuJi Shi and Prof. Sang Hui Im for useful discussion. This work was supported by IBS under the project code, IBS-R018-D1. This work  is supported by NSFC under grants No. 12375088, 12335003 and 12405113.

\bibliographystyle{JHEP}
\bibliography{sample}

\end{document}